\documentclass[sigconf]{acmart} 

  \usepackage{bm}
  \usepackage{subcaption}
  \usepackage{caption}
  \usepackage{hyperref}
  \usepackage{verbatim}
  \usepackage{xcolor}
  \usepackage{listings}
  \usepackage{multirow}
  \usepackage{makecell}
  \usepackage{tabularx}
  \usepackage{comment}
  \usepackage{longtable}
  
  \definecolor{RoyalBlue}{rgb}{0.48, 0.41, 0.93}  

  \newcommand{\eg}{\textit{e.g.,\ }}
  \newcommand{\ie}{\textit{i.e.,\ }}
  \newcommand{\etal}{\textit{et al.\ }}

  \newcolumntype{L}[1]{>{\raggedright\let\newline\\\arraybackslash\hspace{0pt}}m{#1}}
  \newcolumntype{C}[1]{>{\centering\let\newline\\\arraybackslash\hspace{0pt}}m{#1}}
  \newcolumntype{R}[1]{>{\raggedleft\let\newline\\\arraybackslash\hspace{0pt}}m{#1}}

  \newcommand\addauthornote[1]{%
    \if@ACM@anonymous\else
      \g@addto@macro\addresses{\@addauthornotemark{#1}}%
    \fi}

\copyrightyear{2025}
\acmYear{2025}
\setcopyright{cc}
\setcctype{by-nc-nd}
\acmConference[CHI '25]{CHI Conference on Human Factors in Computing Systems}{April 26-May 1, 2025}{Yokohama, Japan}
\acmBooktitle{CHI Conference on Human Factors in Computing Systems (CHI '25), April 26-May 1, 2025, Yokohama, Japan}\acmDOI{10.1145/3706598.3713855}
\acmISBN{979-8-4007-1394-1/25/04}

\begin{document}

\title{Towards AI-driven Sign Language Generation\\ with Non-manual Markers}



\author{Han Zhang}

\orcid{0000-0002-1377-1168}
\authornote{Work done entirely at Apple.}
\affiliation{%
  \institution{University of Washington, USA}
  \country{}
}
\email{micohan@cs.washington.edu}

\author{Rotem Shalev-Arkushin}
\orcid{0009-0009-8376-0171}
\authornotemark[1]
\affiliation{%
  \institution{Tel-Aviv University, Israel}\country{}
}
\email{rotems7@mail.tau.ac.il}

\author{Vasileios Baltatzis}
\orcid{0000-0001-7748-4152}
\affiliation{%
  \institution{Apple, USA} 
  \country{}
}
\email{vbaltatzis@apple.com}

\author{Connor Gillis}
\orcid{0009-0008-3383-2950}
\affiliation{%
  \institution{Apple, USA} \country{}
}
\email{connorgillis@apple.com}

\author{Gierad Laput}
\orcid{0009-0003-6856-2544}
\affiliation{%
  \institution{Apple, USA} \country{}
}
\email{gierad@apple.com}

\author{Raja Kushalnagar}
\orcid{0000-0002-0493-413X}
\authornotemark[1]
\affiliation{%
  \institution{Gallaudet University, USA} \country{}
}
\email{raja.kushalnagar@gallaudet.edu}

\author{Lorna Quandt}
\orcid{0000-0002-0032-1918}
\authornotemark[1]
\affiliation{%
  \institution{Gallaudet University, USA} \country{}
}
\email{lorna.quandt@gallaudet.edu}

\author{Leah Findlater}
\orcid{0000-0002-5619-4452}
\affiliation{%
  \institution{Apple, USA} \country{}
}
\email{lfindlater@apple.com}

\author{Abdelkareem Bedri}
\orcid{0009-0000-6927-901X}
\affiliation{%
  \institution{Apple, USA} \country{}
}
\email{bedri@apple.com}

\author{Colin Lea}
\orcid{0000-0001-7068-3351}
\affiliation{%
  \institution{Apple, USA} \country{}
}
\email{colin\_lea@apple.com}

\renewcommand{\shortauthors}{Zhang \etal}
\renewcommand{\shorttitle}{Towards AI-driven SLG with Non-manual Markers}

\begin{abstract}

Sign languages are essential for the Deaf and Hard-of-Hearing (DHH) community. Sign language generation systems have the potential to support communication by translating from written languages, such as English, into signed videos. However, current systems often fail to meet user needs due to poor translation of grammatical structures, the absence of facial cues and body language, and insufficient visual and motion fidelity. 
We address these challenges by building on recent advances in LLMs and video generation models to translate English sentences into natural-looking AI ASL signers.
The text component of our model extracts information for manual and non-manual components of ASL, which are used to synthesize skeletal pose sequences and corresponding video frames. 
Our findings from a user study with 30 DHH participants and thorough technical evaluations demonstrate significant progress and identify critical areas necessary to meet user needs. 
\end{abstract}

\ccsdesc[500]{Human-centered computing~Ubiquitous and mobile computing design and evaluation methods}
\ccsdesc[500]{Human-centered computing~Accessibility systems and tools}

\keywords{Sign language generation, assistive technology, accessibility, human-centered design, DHH community}
\begin{teaserfigure}
    \centering
    \includegraphics[width=\textwidth]{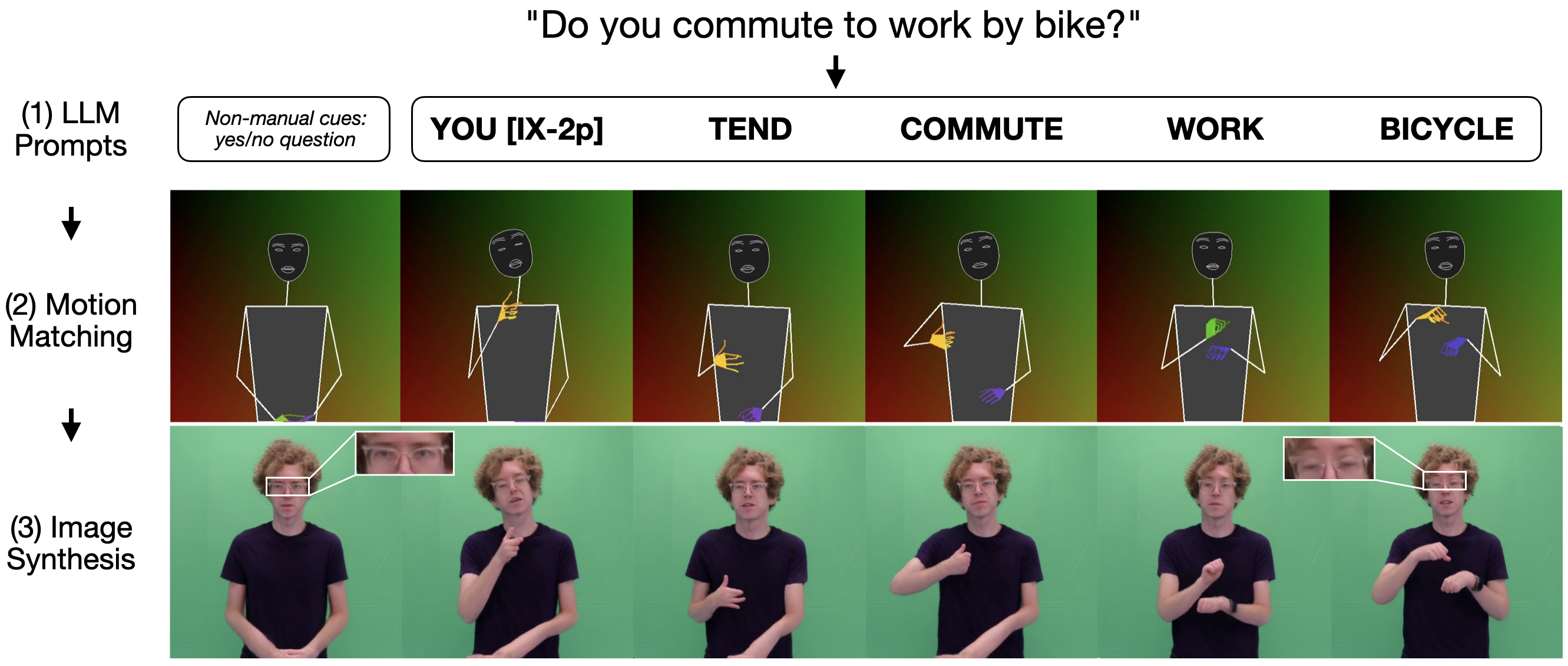}
    \caption{Our prototype translates English text into a photorealistic ASL video which includes both manual and non-manual information. It starts with an English text input (top), and translates it into ASL representations capturing both manual elements (\eg hand movements) and non-manual information (\eg facial expressions). From those, it produces a skeletal pose sequence, and finally converts it into a photorealistic ASL video. In this example, raised eyebrows signal a yes/no question. Without this non-manual marker, the same sentence would be interpreted as a statement.
    \Description[This figure illustrates the process of our prototype system in translating an English sentence  into an American Sign Language video. The top of the figure shows the example English text (``Do you commute to work by bike?''). Below it, the glosses and non-manual markers for each word in the sentence are displayed. The glosses are: IX-2p, TEND, COMMUTE, WORK, and BICYCLE, with a notation indicating a yes/no question. The next row contains a series of five diagrams representing five skeletal poses. Each diagram shows a specific handshape and location corresponding to the glosses and non-manual markers, color-coded for clarity. The bottom row contains images of a person performing the ASL signs, matching the glosses and non-manual markers above. In these images, the person is shown against a green background, demonstrating each sign with corresponding hand positions and facial expressions. Zoomed-in sections highlight the facial expressions associated with the yes/no question. Note that, faces that are not in the zoomed-in sections are blurred to preserve anonymity.]}
    \label{fig:system_overview}
\end{teaserfigure}


\maketitle


\section{Introduction}\label{sec:intro}

Sign languages are crucial for communication within the Deaf and Hard-of-Hearing (DHH) communities~\cite{glickman1993deaf,padden1988deaf}. As naturally-emerging and fully-fledged languages, they enable individuals to convey complex ideas, emotions, and cultural nuances through movements and facial expressions~\cite{emmorey2001language,rastgoo_sign_2021}. Despite their importance for many DHH people, communication barriers between signing and non-signing communities exist due to limited access to skilled sign language interpreters, low levels of sign language proficiency among the general population, and the exclusion of sign languages from most communication technologies designed for spoken or written languages~\cite{mitchell2023many,napier2002sign,bragg_sign_2019}. 

Systems that translate spoken languages into sign languages (sign language generation, SLG) and vice versa (sign language translation, SLT) hold promise to bridge this communication gap~\cite{rastgoo_sign_2021, stoll_text2sign_2020}. In this work we focus on the SLG task, specifically translating from English text to American Sign Language (ASL). Historically, SLG technology has faced criticism from DHH users due to low-fidelity avatars, poor language translation, and oversimplification of sign linguistics~\cite{kipp2011assessing,huenerfauth2009sign,huenerfauth2009linguistically}. Recent research, however, has suggested that improved quality and overcoming technical limitations could increase acceptance among DHH individuals~\cite{quandt2022attitudes, inan2024generating}. Our work uses advances in machine learning (ML) to develop an SLG prototype system and investigate whether technological improvements meet the needs and interests of the DHH and signing community. 

Sign languages combine manual markers---such as hand movements, orientation, and location---with non-manual markers, including facial expressions, head movements, and other body language, to create grammatical structures and convey meaning~\cite{Stokoe1961SignLS,brentari1998prosodic,sandler2006sign}. For example, in ASL, manual markers such as location and movement within the signing space can modify a sign's grammatical function, indicating subjects, objects, or other syntactic roles~\cite{Stokoe1961SignLS}. Similarly, non-manual markers can also indicate critical information, such as a head shake accompanying a sign to denote negation, raised eyebrows and a distinct facial expression to form conditional clauses or emphasize the topic of a sentence or raised eyebrows and a forward head tilt to signal a yes/no question ~\cite{baker1991american,baker1985facial, sandler2006sign}, as exemplified in Figure \ref{fig:system_overview}. Each of these linguistic aspects of ASL presents a challenge for modern SLG systems, given that natural, understandable signing must include sufficient information shown in a fluid manner to convey multiple distinct streams of information.

While recent work on SLG has progressed~\cite{fang2024signllm, hohenberger2002modality, saunders_progressive_2020, moryossef2023open, stoll2018sign, huenerfauth2008generating}, these systems typically take a generic view of signing, often overlooking sign language nuances, including the role of non-manual markers. 
To address these challenges, we prototype a modular ASL generation system designed to produce automated signing by simultaneously focusing on technical improvements, user perceptions, and the unique linguistic structure of ASL. Our system is tailored for open-ended, context-free use cases, allowing users to input an English sentence and generate a signed video that appears natural and comprehensive. Developing an effective SLG system capable of modeling complex signed interactions is a grand challenge that requires interdisciplinary expertise, alongside stewardship from the DHH and signing community~\cite{bragg_sign_2019}. Guided by this principle, our research prototype was developed and refined through collaboration among researchers from diverse fields, including those in computer vision, computer graphics, human-computer interaction, and experts from the DHH and signing communities. It consists of three modules: (1) translating English text into intermediate ASL representations---including English-based glosses to capture manual markers and linguistic information to represent non-manual markers---using few-shot approach with GPT-4o, (2) synthesizing human pose and body motions from these representations using a Motion Matching approach, and (3) generating photorealistic signed video frames representing an ASL signer using an image generation model. 

We conducted both technical evaluations and a user study with 30 DHH signers to assess our prototype system and to gauge the interest of DHH individuals in its use. The technical evaluation examined the translation of English sentences into ASL written representations, including manual and non-manual components, and the generation of signed videos. The user study evaluated translation quality, visual fidelity, and motion naturalness, while gathering perspectives on potential use cases. 
Our findings indicate that the system achieves compelling translation performance relative to reported results in the literature. 
However, there remains significant room for improvement. While participants were frequently able to understand the content of the signed videos, their perceptions on the signing quality, particularly in comparison to real human signers, were less favorable.

In summary, our contributions include: 
\begin{itemize}
    \item In Section \ref{sec:design}, we introduce a modular ASL generation prototype designed to produce natural and comprehensive signed videos that includes non-manual cues.
    \item In Section \ref{sec:technical_eval}, we present technical evaluations of our approach. Results show a BLEU-4 score of $0.276$ for English Text-to-ASL gloss translation, an average precision of 0.91 and recall of 0.97 for detecting non-manual information from English text, and improved video generation performance over baseline methods.
    \item In Section \ref{sec:user_study}, we detail a user study assessing the perceived translation quality of our system, as well as the visual and motion quality of its outputs. Results indicate both potential and tangible areas for improvement, alongside insights into the system's potential use cases (\eg doctor office and video or in-person conversations).
    \item In Section \ref{sec:discussion}, we reflect on our design process, share key insights gained from our design and evaluation process, provide recommendations on how to address remaining challenges, and discuss computational and ethical considerations in the use of our system.
\end{itemize}

While continued effort is needed to advance SLG systems in collaboration with the DHH and signing communities, our work represents an initial step in addressing critical technical challenges and taking a comprehensive approach to ASL. It demonstrates the potential of these systems and encourages further exploration of critical aspects of signing, especially  non-manual markers. 
\section{Background and Related Work}\label{sec:rw}

In this section, we overview Deaf cultures and sign languages, review sign language generation systems, focusing on their technical challenges, and discuss the DHH users' perspectives on sign language technologies. 

\subsection{Deaf Cultures and Sign Languages}\label{subsec:rw_deaf_culture_asl}

In 1970, the term ``Deaf Culture'' was developed to articulate that many Deaf communities possess their own ways of life, characterized by a shared set of values, behaviors, traditions, and goals~\cite{bragg2021fate,ladd2003understanding}. Deaf signing individuals often identify themselves as members of a distinct cultural group~\cite{obasi2008seeing,padden1988deaf}. Among the most treasured aspects of Deaf culture are sign languages, which function both as a mode of communication and a fundamental component of cultural identity~\cite{glickman1993deaf,bragg_sign_2019,bda}. Despite the historical marginalization of sign languages in education and research, approximately 70 million DHH individuals around the world use sign languages, with over 200 different sign languages in use worldwide~\cite{wfd,huenerfauth2009sign,bragg2021fate}. This variability adds to the challenges in creating any sign language technology, in that tools created on the basis of one sign language may not perform well when applied to a different sign language. Across various academic and scientific disciplines, there is a growing consensus that work focusing on sign language is best conducted by groups with linguistic knowledge, alongside authentic cultural knowledge regarding DHH and signing communities~\cite{desai2024systemic,bragg2021fate}.

\subsection{Sign Language Generation Systems}
\label{subsec:rw_SLG}

SLG systems convert written language into signed content. Existing SLG systems typically employ one of two approaches: translating spoken language text directly into pose sequences that represent the corresponding signed translation~\cite{hwang2024universal,hwang2024gloss}, or incorporating an intermediate written representation between the text and pose sequences~\cite{stoll_text2sign_2020, saunders_progressive_2020, moryossef2023open, walsh_sign_2024, xie2024g2p, saunders2020adversarial, arkushin2023ham2pose}. In both cases, the generated pose sequences are ultimately converted into animations of 3D characters~\cite{kim2022sign,kipp2011sign} or photorealistic video using generative computer vision models~\cite{saunders_signing_2022,stoll_text2sign_2020,saunders2020everybody}. 

Research indicates that using an intermediate written representation in SLG systems, preserving linguistic nuances and grammar, results in improved performance~\cite{hwang2024universal, ma2024multi, camgoz_neural_2018}. While graphical systems such as SignWriting~\cite{sutton1974signwriting} and HamNoSys~\cite{hanke2004hamnosys} offer ways to represent signs, they contain only lexical information and do not contain semantic meaning. Consequently, many SLG systems use sign glosses---a written representation of signs using spoken language text (\eg English for ASL glosses) that preserves the meaning and grammatical structure of signs~\cite{liddell2003grammar,desai2024systemic, bragg_sign_2019, muller_considerations_2023}. 

Text-to-gloss translation typically relies on neural machine translation (NMT) models, such as RNNs or Transformers~\cite{stoll_text2sign_2020, stoll2018sign, egea_gomez_syntax-aware_2021, saunders_progressive_2020, walsh_sign_2024, zhu_neural_2023, saunders2022signing}, which require extensive labeled data. To address data limitations, some models incorporate syntax-aware adaptations or data augmentation techniques~\cite{egea_gomez_syntax-aware_2021, zhu_neural_2023}. Nevertheless, alignment with sign language grammar remains a challenge. For example, recent ASL generation systems achieve BLEU-4 scores\footnote{ BLEU-4 is a machine translation metric representing four gram match between prediction and ground truth. A high BLEU-4 indicates strong alignment with the grammar, where BLEU-4 <20\% usually indicate that translations are hard to understand \cite{papineni_bleu_2002}.} of less than 0.002 and 0.124 (on a scale from 0 to 1) for translating English text to ASL glosses~\cite{inan2024generating,zhu_neural_2023}. Recently, large language models (LLMs) trained on extensive corpora have demonstrated state-of-the-art performance in translation tasks, including for low-resource languages, using few-shot prompting~\cite{brown2020language, hendy2023good, peng2023towards}, presenting a promising direction for improving SLG systems. In this work, we adopt one of these state-of-the-art LLMs, achieving a BLEU-4 of 0.276, reflecting a compelling translation performance.

The conversion of glosses into pose sequences is generally approached using either motion models that learn sign representations from sub-sequences of motions~\cite{saunders_progressive_2020, saunders2021mixed, xie2024g2p}, or from a look-up table that stitch and blend pre-recorded sign sequences~\cite{moryossef2023open, stoll_text2sign_2020, stoll2018sign, saunders2022signing, walsh_sign_2024}. The look-up table approach allows producing full signs based on the dictionary, and the main tasks remain selecting context-appropriate sign variants, and generating smooth and natural sign transitions. Techniques for smoothing transitions include motion graphs, smoothing filters, and frame selection networks~\cite{stoll_text2sign_2020, saunders2022signing, moryossef2023open, walsh_sign_2024}. 

The final step, converting pose sequences into videos, remains an active research area focused on achieving natural, realistic, and temporally consistent results~\cite{chan2019everybody, aberman2019deep, liu2019neural, wang2018video, hu2024animate, wang2024disco}. Early methods used generative adversarial networks (GANs) for motion transfer based on pose data~\cite{chan2019everybody, aberman2019deep, liu2019neural, wang2018video}. Following them, SLG works have adapted GANs to generate photorealistic sign videos~\cite{saunders_signing_2022, walsh_sign_2024}. Diffusion models have further advanced image and video generation from pose sequences, showing strong results in generating images and videos~\cite{ramesh2022hierarchical, saharia2022photorealistic, zhang2023adding, huang2023composer, mou2024t2i, hu2024animate, feng2023dreamoving}, hence recent SLG work adapted them for generating avatars from pose sequences~\cite{fang2023signdiff, fang2024signllm}. However, temporal consistency is not always preserved in these videos, and the animated characters may sometimes cause an uncanny feeling among viewers.

Despite these advancements, challenges remain, particularly in handling non-manual markers (\eg eyebrow movements) and achieving high-quality outputs with temporal consistency. One promising method involves learning a dictionary of facial expressions to match each gloss~\cite{walsh_sign_2024}, which enhances visual realism but does not convey additional meaning. This approach often applies the same expression uniformly across sentences, overlooking the contextual nuances of facial expressions and normalizing signers’ faces to face forward, neglecting the subtleties conveyed by directional gaze. Our approach diverges from existing methods by focusing on incorporating non-manual markers while also addressing temporal consistency and enhancing overall visual quality, aiming to create more natural and accurate representation of ASL.

\subsection{User Perspectives on Sign Language Technologies}\label{subsec:rw_user_perspectives}

There is growing recognition that developing effective sign language technologies requires a deep understanding of Deaf culture and sign language linguistics, coupled with the refinement of technical approaches and active involvement of the DHH and signing community throughout the design and implementation process~\cite{prietch_systematic_2022,desai2024systemic,kipp2011assessing}. Collaborative and participatory design approaches that incorporate feedback from DHH individuals are essential for creating culturally appropriate and more widely-accepted technologies~\cite{bragg_sign_2019}.

Historically, sign language technologies have faced high rejection rates within the Deaf community~\cite{Hsu2024unintel,vogel2024factors,gugenheimer2017impact}, largely due to top-down design approaches that lack user feedback and a deep understanding of sign languages~\cite{kipp2011assessing,mohr2017three,zhang2024illuminating,prietch2022systematic}. For instance, wearable sign language translation gloves have been roundly criticized for focusing narrowly on small sets of handshapes while neglecting other essential linguistic elements like facial expressions and torso orientation~\cite{michael2017why}. Additionally, such technologies place the communication access burden on Deaf signers rather than hearing individuals, despite being marketed to improve accessibility for the Deaf community~\cite{visual2019}. In contrast, sign language technologies developed through active involvement with DHH individuals during the design process have generally been more favorably received~\cite{inan2024generating,kipp2011assessing,boudreault2024closed,anindhita2016designing,kipp2011sign}. Moreover, DHH users may value technologies that increase their independence and allow for two-way communication, without reliance on cumbersome physical devices~\cite{michael2017why,hill2020deaf}.

Despite some potential benefits, concerns remain about the accuracy and quality of sign language technologies~\cite{lee2021american,kipp2011sign,huenerfauth2009linguistically,kipp2011assessing,ebling2016building}. A common criticism is that these tools fail to capture the nuances and variations inherent in sign languages, such as personal signing styles and complex grammatical structures, leading to inaccuracies that erode user trust~\cite{kipp2011assessing,lee2021american,kipp2011sign,huenerfauth2009linguistically,ebling2016building}. Historically, technical developments have focused predominantly on single instances of hand shapes while overlooking phrase-level information, facial expressions, and other critical pieces of information~\cite{ebling2016building,huenerfauth2009linguistically,kipp2011assessing}. When it comes to SLG tools, such as signing avatars, these limitations are compounded by additional design challenges. User acceptance is influenced by the visual design and movement of signing avatars and the user interface design more generally. Avatars perceived as robotic or as failing to capture the nuances of human signing can hinder effective communication and result in negative user perception~\cite{kipp2011assessing,tran2023us,huenerfauth2009sign,quandt2022attitudes}. Past work has also recommended reducing the reliance on extensive text-based instructions and offering customizable features, such as for avatar appearance and signing style~\cite{quandt2022attitudes,muir2005perception,tran2023us}. Building on existing literature, this work integrates linguistic and cultural feedback to refine our design choices and improve system performance. 
\section{Sign Language Generation Prototype}\label{sec:design} 

\begin{figure*}[t]
    \centering
    \includegraphics[width=1\linewidth]{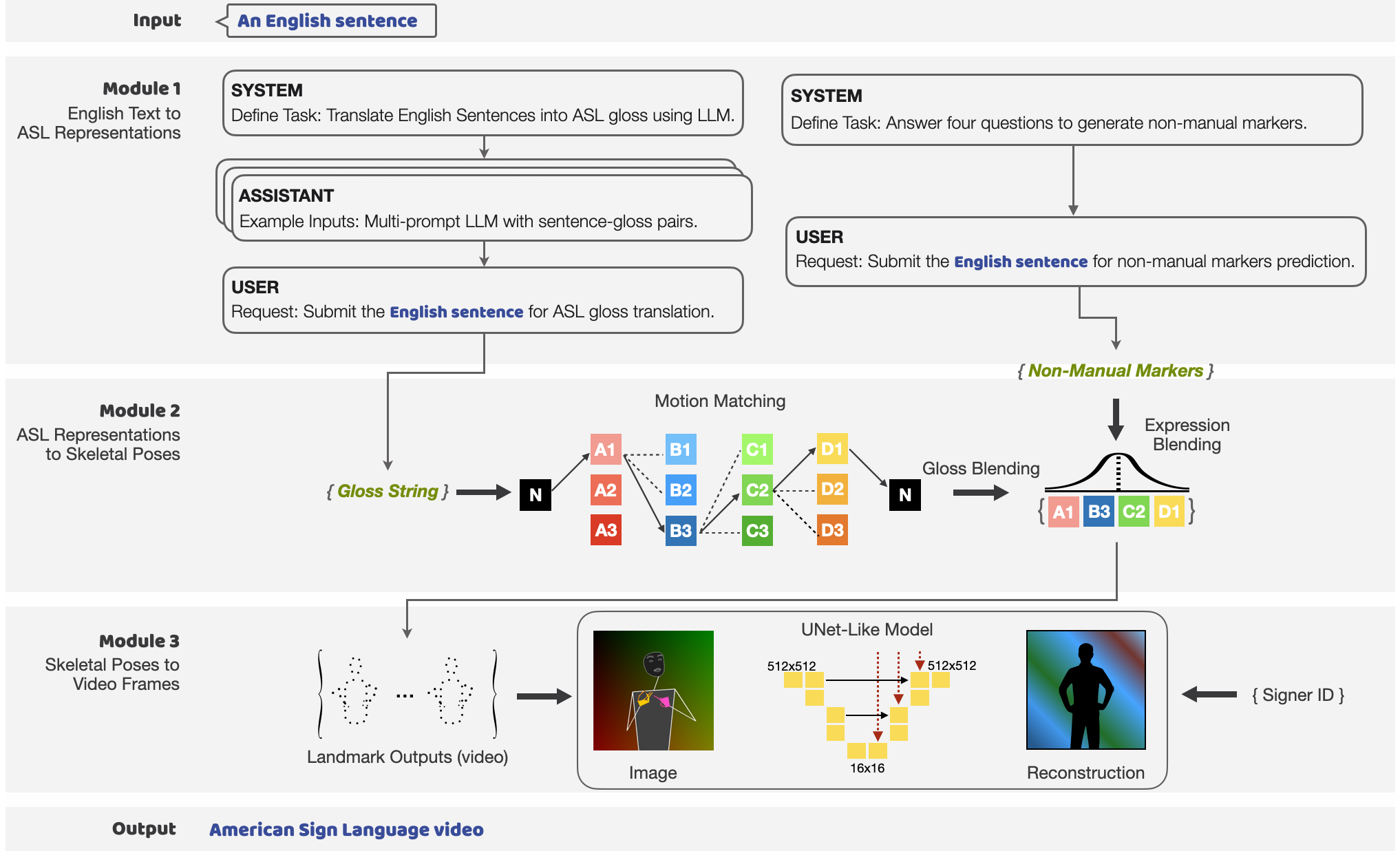}
    \caption{Our prototype includes three self-contained modules. It takes an English sentence as input and generates an ASL video (from top to bottom). Module 1 utilizes a large language model (LLM) to translate the English input into an ASL gloss string and predict non-manual markers. Module 2 employs a Motion Matching approach to generate a skeletal pose sequence from the output of Module 1. Finally, Module 3 uses a UNet-like model, which given an individual signer's appearance and style (Signer ID), transforms the skeletal pose sequence into signing frames. These are then combined to produce the final signing video. 
    \Description[The figure depicts the architecture of our system for generating American Sign Language (ASL) videos from English text, divided into three modules. The first module translates English sentences into ASL glosses and predicts non-manual markers like facial expressions and head movements using a large language model and specific questions. The second module converts the gloss strings into skeletal poses through motion matching and blending techniques to integrate both manual and non-manual components. The third module uses a U-Net-like model to transform these skeletal poses into video frames, refining the visual details based on the signer’s appearance and style. The final output is a video of a photorealsitic AI signer performing ASL, reflecting the original English input.]
    }
    \label{fig:system_arch}
\end{figure*}

In this section we describe our SLG prototype, which generates ASL videos with manual markers---such as hand shape, location, movements, and palm orientation---as well as non-manual markers, including facial expressions and eyebrow movements. 
Our focus is on context-free settings, where each sentence is translated independently. We used a modular approach for our system design (Figure \ref{fig:system_arch}), allowing increased flexibility and interpretability of each module.

The prototype consists of three components: \textbf{Module 1: English Text to ASL Representations}, which leverages a Large Language Model (GPT-4o~\cite{achiam2023gpt}) to translate an English sentence into English-based ASL glosses and to detect linguistic information relevant to non-manual markers; \textbf{Module 2: ASL Representations to Skeletal Pose Sequence}, which takes the LLM outputs and employs a Motion Matching approach to synthesize a skeletal pose sequence; and \textbf{Module 3: Skeletal Pose Sequence to ASL Signed Video}, which generates signed video frames representing a photorealistic ASL signer. 
This modular approach allows for future improvement of the system as the technology advances, by allowing each part to be changed separately. This prototype was iteratively refined within the research team. Insights from these researchers and other collaborators fluent in ASL helped to guide improvements in translation quality, visual and motion quality, and information conveyance.

\subsection{Module 1: English Text to ASL Representations}\label{sebsec:module1}

We used an enhanced gloss-based approach that translates an English sentence into an intermediate ASL gloss, including both manual and non-manual information, which is then utilized by subsequent modules. Given the ability of LLMs to naturally absorb and generate grammatical rules, structures, and nuances~\cite{brown2020language,radford2019language}, we used GPT-4o\footnote{Specifically, we used the model gpt-4o-2024-05-13.}~\cite{achiam2023gpt}, a state-of-the-art LLM, to perform two key tasks: (1) translate an English sentence into English-based glosses and (2) detect if the English sentence contains linguistic features associated with specific facial expressions (Module 1 in Figure~\ref{fig:system_arch}). GPT-4o was selected based on our preliminary experiments with various versions of the GPT models. Detailed experimental results are presented in Appendix \ref{appendix:llm_experiments}. 

For the first task, we adopted a prompting-based approach using LLMs with ``in-context learning''~\cite{xu2024misconfidence}, inspired by recent work on low-resource machine translation~\cite{guo2024teaching}, where dataset sizes are too small to train large-scale translation models. This approach allows the model to adapt and perform specific tasks by interpreting examples or instructions directly embedded in the input text, without requiring explicit retraining~\cite{brown2020language}. To improve performance, we added 1,494 in-context examples of English sentence-gloss pairs to our prompt from the ASLLRP dataset (representing 80\% of the dataset). Given the limited window of GPT-4o (\ie 128,000 input tokens), which restrict the number of examples that can be included in a single prompt, we used a ``multi-prompting'' approach. This method involved splitting the examples into multiple batches and iteratively prompting GPT-4o with each batch. In addition, we asked the LLM to constrain its output by generating glosses within the vocabulary established by our text-to-gloss dictionary described below.

For the second task, we adopted a zero-shot prompting approach, asking the model to predict linguistic features associated with specific facial expressions without any in-context examples. The idea of linguistic predictions was inspired by prior research suggesting that non-manual expressions corresponding to specific grammatical markers, such as raised eyebrows or head tilts, typically involve a consistent set of behaviors that convey meaning within sign language~\cite{neidle2002signstream,baker1983microanalysis}. In this work, we focus primarily on eyebrow movements. To this end, we asked the model to predict whether a given English sentence: is (1) a yes-no question, (2) a wh-question, (3) a conditional statement, and/or (4) contains negation. The outputs from both tasks are then used to generate skeletal poses that are compatible with the subsequent modules, enhancing the integration of non-manual markers. 

This approach addresses two common limitations of gloss-based ASL representations: (1) their tendency to deviate from ASL grammar, and (2) their inability to fully capture the context and expressiveness necessary for conveying the full semantics of a sentiment. 

\paragraph{Dataset and Implementation Details} After reviewing the available ASL datasets (see Appendix ~\ref{appendix:ASL} for more details), we selected the ASLLRP~\cite{neidle_asl_2022} dataset for Module 1. The ASLLRP dataset contains continuous sentence-level ASL videos, isolated ASL videos, ASL glosses, and corresponding English translations. This dataset provides detailed annotations, including textual annotations (\eg English-based glosses for lexical signs, fingerspelling, classifiers, name signs, and gestures), manual markers (\eg number of hands used, alternating hand movements), and non-manual markers (\eg head position and movements, eye gaze, and mouth movements). 

\paragraph{Data Preprocessing} While ASLLRP provides the most comprehensive information required for our task, the data is dispersed across various resources and editions. To make effective use of this dataset, we first consolidated these disparate resources into a unified framework, extracting 2,119 English sentence-gloss pairs along with their corresponding signing videos. The signing videos were then trimmed to isolate specific sign language utterances for our subsequent tasks. To minimize translation errors, we removed gloss annotations that did not alter the overall meaning of the sentence when omitted and standardized all glosses related to fingerspelling. All these changes were done by consulting team members fluent in ASL. We also excluded glosses for classifiers due to their limited sample sizes. After data cleaning, we retained 1,843 English sentence-gloss pairs. Next, we developed a word-gloss dictionary to improve consistency in sign representations of words across different sentences, resulting in 3,915 word-gloss pairs. For the 43 out-of-vocabulary (OOV) words that lacked corresponding videos, we employed fingerspelling as an alternative representation. Finally, four of our researchers conducted a ground truth correction to resolve misalignments between the linguistic labels for the four types of non-manual information and the English text, ensuring the labels more accurately reflected the text content. A more detailed description of our data preprocessing process can be found in Appendix \ref{appendix:data_prep}. The conventions used for re-annotating the glosses in this work are summarized in Table \ref{tab:gloss_convention}.

\begin{figure*}[t]
    \centering
    \includegraphics[width=1\linewidth]{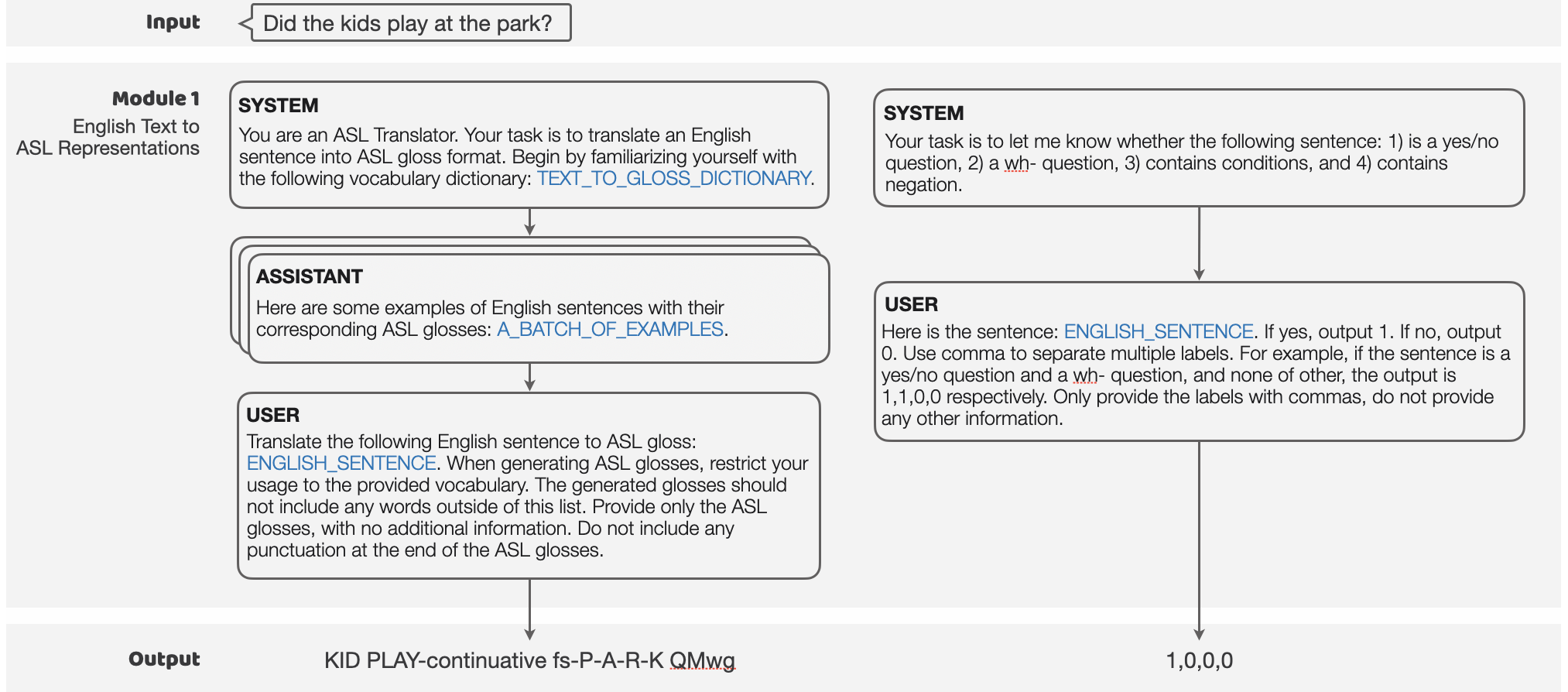}
    \caption{An example from Module 1 showcases two tasks: on the left, translating an English sentence into its corresponding ASL gloss, and on the right, predicting the linguistic features of the same English sentence. \textcolor{RoyalBlue}{\textsf{TEXT$\_$TO$\_$GLOSS$\_$DICTIONARY}} represents the examples provided to the LLM for each shot. \textcolor{RoyalBlue}{\textsf{A$\_$BATCH$\_$OF$\_$EXAMPLES}} refers to the examples we provide to the LLM each shot. \textcolor{RoyalBlue}{\textsf{ENGLISH$\_$SENTENCE}} indicates the user-provided input, which, in this example, ``Did the kids play at the park?''
    \Description[This figure uses an example to illustrate a flowchart describing the steps in our Module 1. The process starts with an input sentence at the top: ``Did the kids play at the park?'' Then the Module 1: English Text to ASL Representations is depicted, where two sub-process occur. On the left side, the task is to let the large language model to translate the given English sentence to the corresponding glosses. Specifically, we prompt the system to act as an ASL translator, by providing the following prompt: You are an ASL translator. Your task is to translate an English sentence in o ASL gloss format. Begin by familiarizing yourself with the following vocabulary dictionary: TEXT underscore TO underscore GLOSS underscore DICTIONARY. Then we provide the model with in-context example, by providing the following prompt: Here are some examples of English sentences with their corresponding ASL glosses: A underscore BATCH underscore OF underscore EXAMPLES. Lastly, we ask the model to translate the given English sentence into ASL glosses, while restricting the translation to our word-to-gloss dictionary as its vocabulary. The prompt we provide to the model as the user is: Translate the following English sentence to ASL gloss: ENGLISH underscore SENTENCE. When generating ASL glosses, restrict your usage to the provided vocabulary. The generated glosses should not include any words outside of this list. Provide only the ASL punctuation at the end of the ASL glosses. On the right side, the task is to predict the linguistic features regarding the non-manual markers of the given English sentence. Specifically, we prompt the system with the following information: Your task it to let me know whether the following sentence: 1) is a yes/no question, 2) is a wh- question, 3) contains conditions, and 4) contains negation. The user instructs this process to label each feature as 1 (yes) or 0 (no). The user prompt is: Here is the sentence: ENGLISH underscore SENTENCE. If yes, output 1. If no, output 0. User comma to separate multiple labels. For example, fi the sentence is a yes/no question and a wh- question, and none of other, the output is 1,1,0,0 respectively. Only provide the labels with commas, do not provide any other information. At the bottom of this flowchart, the output section shows the results of translating the input sentence into ``KID PLAY-continuative fs-P-A-R-K-QMwg.'' The sentence type labels are outputted as 1,0,0,0, indicating the sentence is a yes/no question.]}
    \label{fig:module_1}
\end{figure*}

\paragraph{LLM Translation and Classification} 
We used few-shot and zero-shot prompting over GPT-4o \cite{achiam2023gpt} to perform these tasks.
Our prompts were designed to ensure the outputs could be directly used 
for downstream tasks and systematic evaluation. 
Figure \ref{fig:module_1} overviews the process and prompts, and shows a usage example. More examples of prompts can be found in Table \ref{tab:prompt_engineering}. For the translation task, we structured the process by first defining the task for the system. Next, we provided the model with context using English word-gloss pair examples for few-shot learning. Finally, we asked the model to translate each English sentence into ASL glosses, while restricting the translation to our word-gloss dictionary as its vocabulary. For the linguistic features task, we also started by defining the task for the system, and then, using zero-shot prompting, asked the model to classify the linguistic features in the English sentence, \ie if it contains a yes/no question, a wh-question, a condition, and/or a negation. Zero-shot was enough in this case, because GPT was extensively trained over English text. The exact prompts used for this process are shown in Figure \ref{fig:module_1}.

\subsection{Module 2: ASL Representations to Skeletal Pose Sequence}\label{subsubsec:module2} 
The goal of this module is to take the gloss and non-manual LLM outputs and generate a sequence of skeletal poses at video frame rate, which expresses the input English phrase. We based our approach on Motion Matching, a widely used technique in the Computer Graphics community~\cite{buttner2015motion,clavet2016motion,holden2020learned}, which takes a large dictionary of short character animations and an input signal and intelligently blends clips together to form a cohesive video. 
Given the gloss input, a sequence of reference clips is chosen from the dictionary using an optimization function that minimizes the signing concept of ``economy of motion.'' This principle prioritizes the ``best'' sign by minimizing the distance between the body position at the end of the previous sign and the start of the next. The selected clips are then linearly blended together to create a cohesive sequence. 
The non-manual predictions are used as input to an expression blending part of the model which takes the glossed output and augments the facial expressions, in particular targeting eyebrow motion. 
Our signing dictionary derived from ASLLRP contains 12,681 signed pose sequences, with many repetitions of each sign, which are labeled with the 3,915 glosses noted above. 

Motion Matching typically comprises of three components: (1) a definition for how we represent pose sequences and how they are used for generating the pose sequence dictionary, (2) similarity and optimization functions for identifying the ``best'' elements for a sequence, and (3) a blending function to create the resulting pose sequence. 
See Figure \ref{fig:system_arch} (Module 2) for a visual description. The first step chooses and blends the best sign variants. A second step applies expression blending, which augments the pose sequences with non-manual markers to refine facial expressions. 

\paragraph{Skeletal Pose Representation \& Sign Dictionary}
Whole body, face, and hand skeletal keypoints are extracted from all isolated sign videos in ASLLRP using Mediapipe~\cite{lugaresi2019mediapipe}, using 3D information for hands and 2D information for the others.
We preprocessed this data in three ways. First, we imputed keypoints that were missing due to occlusion issues and poor tracking. For missing keypoints at the beginning or end of a sequence, we filled in points with neutral poses where the hands were positioned together just below the viewpoint from the camera. All other missing keypoints were linearly interpolated using valid keypoints from timesteps before and after within that sequence. One exception was with fingerspelling, where we intentionally kept the non-dominant hand in the same neutral position to avoid jumps between letters in a word. Second, we normalized all keypoints in space so that position and scale of the body and head were consistent across sequences. This alleviated differences in camera position between videos and body shapes between signers. For positioning, we relied on the first frame of each sign with the average shoulder position in subsequent frames relative to that first frame. Lastly, we trimmed the start and end of each sign using annotations from the ASLLRP dataset. For fingerspelling, we sped up the clips to account for the discrepancy between the slower performance in the isolated sign video clips and the faster pace typically used in-situ \cite{quinto2010rates}. 

\paragraph{Optimization functions}
There are many different ways to articulate the same sign for emphasis, style, and convenience~\cite{baker1991american,brentari1998prosodic}. For most signs in our dictionary we have multiple examples of each sign. Often these variants convey the same meaning, but are performed by different signers. Sometimes the meaning does vary. For example, ``big'' might have versions that convey a medium-big size and a large-big size or a sign might be shown using newer and antiquated styles. In short of having sufficient linguistic information to differentiate sign variations, we select sign variants based on minimizing movement rather than incorporating other linguistic factors. In the signing community this is sometimes referred to as minimizing the ``economy of motion,'' where an individual may blend together sign variations based on which is physically more efficient.  Mathematically, given a vector of keypoint locations $x_{i,t}^p$ where $i$ is a valid gloss index, $t$ is a frame index within a clip, and $p$ is a body part (body, face, hands), we compared the Euclidian distance using a weighted average of the current gloss $i$ and a candidate subsequent gloss indexed by $j$:
\begin{equation}
	d(i, j) = \sum_{p \in \{body,\ face,\ hands\}} \alpha_{p} \cdot \left\|x_{i,T}^p - x_{j,0}^p\right\|_2^2,
\end{equation}
where $\alpha_p$ is a weighting value for each body part, and $T$ is the final frame in the clip. Values of $alpha$ were chosen to prioritize importance of the body and prevent large changes in posture.

The final sequence of sign videos was determined by minimizing the differences (maximizing the similarity) across all glosses output from the LLM. This was achieved by a greedy algorithm that selected sign videos with the correct gloss labels, prioritizing those where the beginning of the clip was most similar to the end of the previous clip. 

\paragraph{Gloss \& Expression Blending}
We generated a preliminary pose sequence by linearly blending together the start and end of the pose sequence from each chosen gloss instance, using the first and last 20 frames of each clip (at 90 Hz). To increase smooth transitions, we appended half-second neutral pose to the beginning and end of each sequence of videos, which was also interpolated with the gloss videos. We then used the predicted non-manual marker information to augment the facial expressions holistically after stitching the videos together. Specifically, we adjusted the position of the eyebrows throughout the video to reflect whether a sentence was a yes/no question, wh-question, or neither. The output of this module is a sequence with body, face, and hand keypoint poses for a full video. 

\subsection{Module 3: Skeletal Poses to Video Frames}\label{subsec:module3}
The last module converts the generated pose sequences into a sequence of photorealistic images. 
First, the input 2/3D skeleton poses are rasterized by drawing the skeletal positions onto an image. 
Second, these skeletal images are used as input to an image-to-image neural network, which outputs photorealistic images. We choose to generate videos that resemble ``live'' signers, in an effort to mitigate confounds that could arise in accurately representing signs with more stylized avatars.

The design decisions regarding the rasterization function---the way the skeleton is drawn---play a critical role in the performance of the image-to-image model. In the baseline rasterization function used by previous work~\cite{zhang2023adding, hu2024animate}, each landmark position was represented by a circle on a 2D image with a monochromatic (black) background, with straight lines connecting the hand and torso landmarks. This is consistent with the commonly used drawing functions within the Mediapipe~\cite{lugaresi2019mediapipe} library. In contrast, in our drawing function, instead of scattered, connected circles, each body part (\ie hands, body, face) was represented as a convex polygon, with additional connections drawn between the face and the entire body. Each body surface was drawn with a different shade of gray and each hand uses a different color palette where each finger is a different shade. We use the 3D data from each hand to determine the palm orientation (``in'' versus ``out'') using the surface normal of landmarks surrounding the palm and augment hand colors based on this orientation. Moreover, the background in our dataset varies per person and we find that using a rasterized background color with shades of black-to-red going from top to bottom and black-to-green going from left to right improves the stability of the generated images. The proposed rasterization function significantly improves the image quality and background stability with emphasis to differentiating the hands and individual fingers, disambiguating occlusions originating in overlap between body parts, and differences in the backgrounds of each image.

Although there have been large advances in photorealistic image generation of humans using diffusion models (\eg ControlNet~\cite{zhang2023adding}), results tend to lack temporal consistency and often do not represent hands accurately. Hence, our work builds on image-to-image translation models~\cite{isola2017image, brooks2023instructpix2pix, tumanyan2023plug, hertz2022prompt, zhang2023adding}, while adding modern architectures and loss functions. 
Specifically, we used a U-Net architecture ~\cite{ronneberger2015u}, with the encoder and decoder backbones using neural building blocks from the architecture in Imagen~\cite{saharia2022photorealistic}. 
Unlike Imagen, which uses text as an additional input to the system, we condition the decoder using the signers' identity. 
This is especially important because we use data from many different Signer IDs as part of the same model.
This enables the network to output different visual appearances for each Signer ID in the dataset, which is used when training the network and at inference time. 

The model is trained with a combination of three losses. These are an L1 term between the entire generated output frame and the target input frame, an L1 term only on the hand region, and an LPIPS term~\cite{zhang2018unreasonable}, which is a learned metric that measures perceptual similarity between the output frame and the target input frame. 
The total loss used to train the model is the sum of the whole frame L1 loss, the hand-specific L1 loss, and the perceptual loss.

\paragraph{Dataset and Implementation Details} 

Our primary dataset for image generation experiments is How2Sign~\cite{duarte_how2sign_2021}\footnote{There is ambiguity as to which individuals in How2Sign gave permission to use their likeness in publications. Thus, for visualization purposes within this paper and supplemental material, we trained additional models that contain the identity of two other people who have given their permission. Qualitatively, these results are representative of the How2Sign results.}.
Although it doesn't contain glosses, in contrast to ASLLRP, which has varied quality across videos, How2Sign contains 35K high-resolution clips of ASL with a vocabulary size of over 16K word tokens. The high resolution and overall data quality of How2Sign helps the model to learn fine-grained and high-quality visual representations of ASL. 

While the overall image quality is generally high, there are problems with skeleton tracking, especially when there is significant motion blur or there is ambiguity in hand pose. 
Thus, when training the model, we discard lower quality frames
in efforts to learn more precise mappings between skeletons and photo-realistic humans.
We accomplish this by performing automated visual checks in both image and skeletal pose spaces. 
In image space, we use optical flow to detect motion blur by analyzing the flow vectors between two consecutive frames using Farneback's method ~\cite{farneback2003two}. 
In pose space, we check for sudden large changes in landmark positions between consecutive frames, which might indicate inaccuracies due to motion blur. 
Specifically, we compared the current landmark positions to the mean landmark positions over a sliding window of predefined size. While signing, hands tend to move more than the body, so the pose conditions are imposed only for each hand instead of including the entire body and face.

\section{Technical Evaluation}\label{sec:technical_eval}

We conducted technical evaluations to assess the performance of our proposed system in translating English text into intermediate ASL representations and generating signed videos. The following sections provide a detailed account of each evaluation, including the experimental procedures and the corresponding evaluation results. Note that a direct quantitative comparison with previous work is challenging due to the use of different datasets~\cite{inan2024generating,zhu_neural_2023,moryossef_data_2021}, output modalities, or gloss-less approaches~\cite{baltatzis2024neural}. For instance, while benchmarks for English text-to-ASL gloss translation often use datasets from other languages, benchmarks specific to ASL gloss translation are lacking. Additionally, for video generation, \cite{baltatzis2024neural} employs the How2Sign dataset and produces SMPL-X 3D human body model poses, whereas our system generates photorealistic videos. These differences in output (3D models vs. photorealistic videos) and their end-to-end design, which precludes comparison of intermediate components, make direct comparisons impractical.

\subsection{English Text to ASL Represenetations}\label{subsec:technical_eval_exp1} 

\subsubsection{English Text-to-ASL Gloss Translation}\label{subsubsec:english_to_gloss}

We conducted ablation studies to determine the optimal model configuration for translating English sentences into English-based glosses (as illustrated on the left side of Module 1 in Figure \ref{fig:system_arch}). Specifically, we examined four key factors: the impact of data preprocessing, the number of in-context examples fed to GPT, the effectiveness of generating glosses within the vocabulary established in our word-to-gloss dictionary, and the necessity of guiding GPT to learn ASL grammar rules\footnote{The ASL grammar rules we provided to GPT-4o can be found in \ref{asl_grammar_rules} in Appendix.}. For the number of English-to-gloss examples, we experimented with 600 (33\% of dataset) and 1,474 (80\% of dataset) sentences from ASLLRP. The dataset was randomly split into a 80/20 ratio to mitigate inconsistencies in distribution. We report BLEU~\cite{papineni_bleu_2002} scores (1 to 4 grams) and ROUGE-L~\cite{lin_rouge_2004} scores, two widely used metrics in the machine translation community~\cite{baltatzis2024neural,saunders_progressive_2020,saunders2020adversarial,fang2024signllm}. Additionally, for a more comprehensive evaluation, we include METEOR~\cite{banerjee_meteor_2005}, CHrF~\cite{popovic_chrf_2015}, TER~\cite{snover_study_2006}, and SacreBLEU~\cite{post2018call}, which are also commonly applied in the literature to assess text-to-gloss translation quality~\cite{egea_gomez_syntax-aware_2021,zhu_neural_2023,forster_extensions_nodate}.
\begin{table*}[t]
\caption{Evaluation results of translating English text to ASL glosses (Task on the left side in Module 1). ``Prep.'' denotes Preprocessing. \bm{$^*$}All BLEU-4 and SacreBLEU scores are identical. \bm{$\uparrow$} indicates that higher values represent better performance, while \bm{$\downarrow$} indicates that lower values represent better performance. Best results in \textbf{bold}. Note: If ``Data Prep.'' is set to ``No'', the model was not restricted to generating glosses within the word-to-gloss dictionary vocabulary, as the dictionary generation is part of our preprocessing step.
\Description[This table presents evaluation results of translating English text to English-based glosses. The first row contains twelve headers, including Data preparation, number of examples, limited vocabulary, grammar rules, bleu-1, bleu-2, bleu-3, bleu-4, rouge-l, meteor, chrf, and ter. Regarding the columns, the first column indicates whether data preparation is applied (yes or no). The second column, number of examples, shows two sets: 600 and 1474 examples, with specific portions of the entire dataset (33\% or 80\%). The third column, limited vocab, indicates whether a limited vocabulary is use (yes or no). The fourth column, grammar rules, shows whether grammar rules are applied (yes or no). The remaining columns display various evaluation metrics with the symbol uparrow indicating higher values are better, and downarrow indicating lower values are better. Key findings are: first, when data preparation is yes, 600 examples, with no limited vocabulary and grammar rules applied, bleu score (1 to 4 grams) range from 0.470 to 0.214, with the best bleu-4 score at 0.214; second, with 1474 examples and the same conditions, the bleu scores improve, reaching a bleu-4 score of 0.276; third, metrics such as rouge-l, meteor, and chrf show similar trends, generally improving with more examples and grammar rules; fourth, the best scores are bolded across all metrics, indicating the optimal settings for this translation model.]}\label{tab:text-to-gloss_eval_results}
\renewcommand{\arraystretch}{1.1}
 \resizebox{\textwidth}{!}{
\begin{tabular}{L{0.04\textwidth}|C{0.13\textwidth}|C{0.06\textwidth}|C{0.08\textwidth}| R{0.08\textwidth} R{0.08\textwidth} R{0.08\textwidth} R{0.09\textwidth} R{0.1\textwidth} R{0.09\textwidth} R{0.06\textwidth} R{0.05\textwidth}}
\toprule\hline
\multicolumn{1}{c|}{{\textbf{\makecell[c]{Data \\ Prep.}}}} & \multicolumn{1}{c|}{{\textbf{\makecell[c]{Number of \\ Examples}}}} &\multicolumn{1}{c|}{{\textbf{\makecell[c]{Limited \\ Vocab}}}}&\multicolumn{1}{c|}{{\textbf{\makecell[c]{Grammar \\ Rules}}}} & \multicolumn{1}{c}{{\textbf{\makecell[c]{BLEU-1 \bm{$\uparrow$}}}}} &\multicolumn{1}{c}{{\textbf{\makecell[c]{BLEU-2 \bm{$\uparrow$}}}}} & \multicolumn{1}{c}{{\textbf{\makecell[c]{BLEU-3 \bm{$\uparrow$}}}}}
& \multicolumn{1}{c}{{\textbf{\makecell[c]{BLEU-4\bm{$^*$} \bm{$\uparrow$}}}}} 
& \multicolumn{1}{c}{{\textbf{\makecell[c]{ROUGE-L \bm{$\uparrow$}}}}} 
& \multicolumn{1}{c}{{\textbf{\makecell[c]{METEOR \bm{$\uparrow$}}}}} 
& \multicolumn{1}{c}{{\textbf{\makecell[c]{CHrF \bm{$\uparrow$}}}}} & \multicolumn{1}{c}{{\textbf{\makecell[c]{TER \bm{$\downarrow$}}}}} \\\hline 
\multirow{4}{*}{No}& \multirow{2}{*}{600}& \multirow{2}{*}{-} & Yes& 0.295 & 0.204 & 0.151 & 0.116 & 0.573& 0.352 & 0.426 & 0.668\\
 & & & No & 0.358 & 0.260& 0.201 & 0.158 & 0.591& 0.386 & 0.454 & 0.644\\\cline{2-12}
 & \multirow{2}{*}{1,474}& \multirow{2}{*}{-} & Yes& 0.379 & 0.280& 0.220& 0.177 & 0.603& 0.406 & 0.462 & 0.625\\
 & && No & 0.404 & 0.303 & 0.239 & 0.192 & 0.611& 0.432 & 0.472 & 0.619\\\hline
\multirow{8}{*}{Yes} & \multirow{4}{*}{\makecell[c]{600 \\(33\% of the \\ entire dataset)}}& \multirow{2}{*}{No}& Yes& 0.470& 0.336 & 0.255 & 0.197 & 0.617& 0.498 & 0.487 & 0.585\\
 & && No & 0.487 & 0.355 & 0.273 & 0.214 & 0.627& 0.502 & 0.503 & 0.572\\\cline{3-12}
 & & \multirow{2}{*}{Yes} & Yes& 0.520& 0.390& 0.305 & 0.241 & 0.641& 0.530 & 0.522 & 0.556\\
 & && No & 0.520 & 0.387 & 0.302 & 0.237 & 0.642& 0.523 & 0.528 & 0.554\\\cline{2-12}
 & \multirow{4}{*}{\makecell[c]{1,474 \\ (80\% of the \\ entire dataset)}}& \multirow{2}{*}{No}& Yes& 0.501 & 0.378 & 0.298 & 0.239 & 0.646& 0.534 & 0.521 & 0.537\\
 & && No & 0.513 & 0.386 & 0.303 & 0.243 & 0.645& 0.532 & 0.519 & 0.548\\\cline{3-12}
 & & \multirow{2}{*}{Yes} & Yes& 0.545 & 0.415 & 0.329 & 0.265 & 0.662& 0.551 & 0.544 & \textbf{0.524}\\
 & && No & \textbf{0.556} & \textbf{0.427} & \textbf{0.341} & \textbf{0.276} & \textbf{0.664} & \textbf{0.560} & \textbf{0.549} & 0.526 \\\hline
 \bottomrule
\end{tabular}}
\end{table*}

As shown in Table~\ref{tab:text-to-gloss_eval_results}, our ablation study results indicate that data preprocessing improves the LLM's performance in translating English text to English-based glosses. Similarly, providing the LLM with more examples, when they are chosen randomly, and limiting the generated glosses to those within the word-to-gloss dictionary results in higher BLEU (1 to 4 grams), ROUGE-L, METEOR, and CHrF scores, along with lower TER scores, all of which suggest enhanced model performance. 


\begin{figure}
    \centering
    {\includegraphics[scale=0.36]{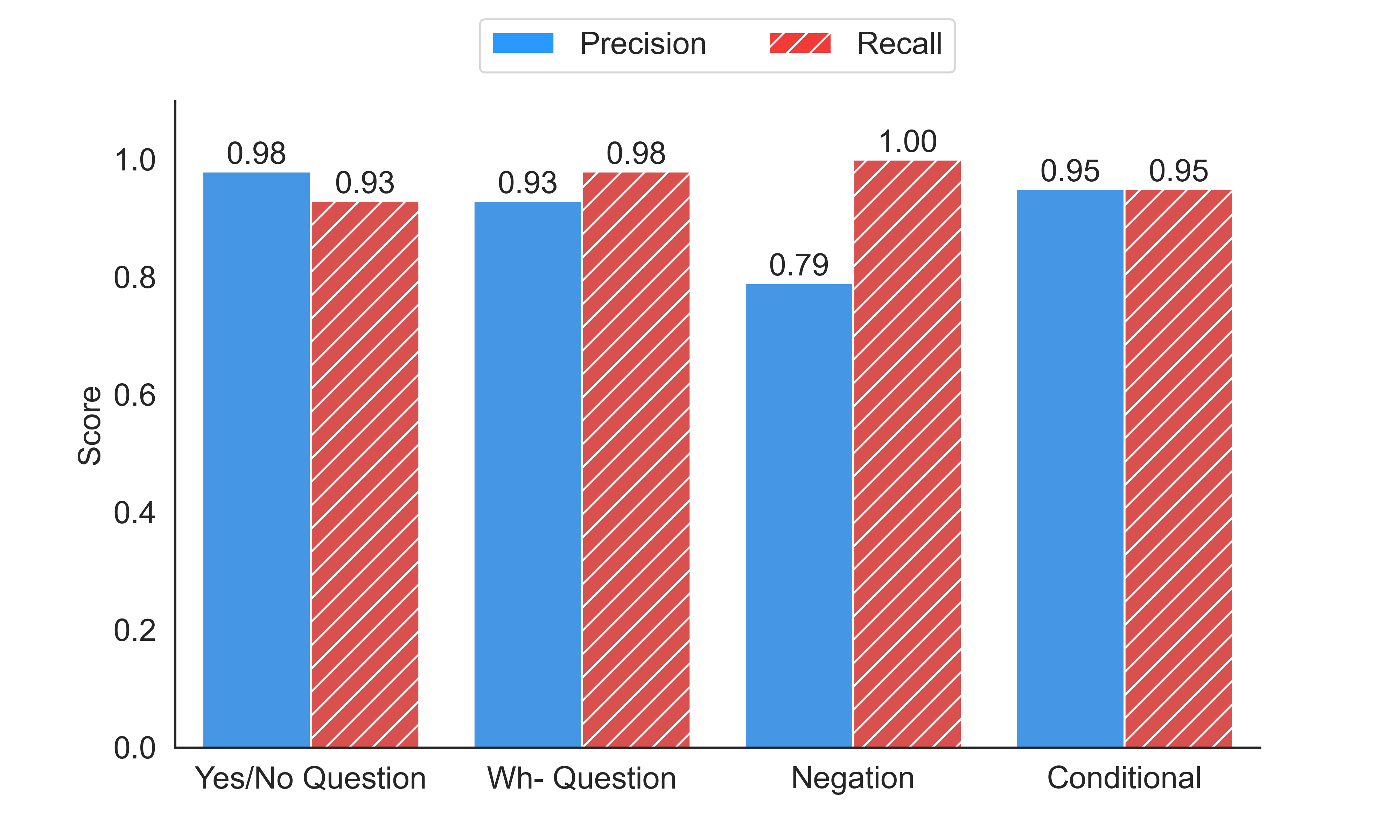}}
    \caption{Model performance in detecting linguistic features to generate non-manual marker information. The model demonstrates high performance across all categories, with particularly high recall for detecting negation and precision for detecting yes/no questions. Precision for negation, however, is relatively lower at 0.79.
    \Description[This grouped bar chart displays the model's precision and recall performance across four types of linguistic features: yes/no question, wh- question, negation, and conditional (from left to right). The blue bars reprent precision, while the red bars represent recall. From left to right, the precision and recall for each type is 0.98 and 0.93 for yes/no question, 0.93 and 0.98 for wh- question, 0.79 and 1.00 for negation, and 0.95 and 0.05 for conditional.]}
    \label{fig:non_manual_results}
\end{figure} 

Interestingly, most experiments showed that adding grammar rules did not improve the model’s translation ability, however, there were some exceptions. 
For example, when data preprocessing was applied and the LLM was provided with 80\% of the entire dataset without limiting the generated glosses to the word-to-gloss vocabulary, we observed mixed results. Specifically, BLEU (1 to 4 grams) scores suggested better model performance without adding grammar rules to the LLM, while other metrics indicated the opposite trend. Furthermore, although direct comparisons are challenging, our system demonstrates compelling translation performance compared to existing results reported in the literature, achieving a BLEU-4 score improvement from 0.191 to 0.276. Table \ref{tab:text-to-gloss_sota} in Appendix \ref{appendix:existing_text-to-gloss_results} summarizes the existing English Text-to-ASL gloss translation results reported in the literature.

\subsubsection{Linguistic Predictions}\label{subsubsec:linguistic_predictions}

Falsely predicting linguistic features for a sentence could result in unnecessary non-manual markers added to the sequential poses and video frames, potentially leading to confusion in the generated ASL videos. To evaluate the performance of GPT-4o in detecting linguistic features regarding the four questions---whether a sentence is a yes/no question, wh-question, conditional statement, and/or contains negation---we calculate precision and recall for each type of prediction. 

Figure \ref{fig:non_manual_results} summarizes the model’s performance in detecting linguistic features within a given English sentence across the four conditions. Overall, the model demonstrates high accuracy across these tasks, particularly in identifying questions and conditional statements. The relatively low precision for negation (precision=0.79) suggests that the model occasionally incorrectly identified negation in sentences where the human-labeled ground truth did not indicate negation presence. We analyzed these cases and discovered that in most, the sentences include negative sentiment, \eg "Why do you hate video games?" or "My sister blamed me but I am innocent!"

\subsection{Video Generation}
\label{subsec:technical_eval_exp3}

We evaluated our system's performance in generating signed videos (Modules 2 and 3) using quantitative metrics commonly used for human video generation~\cite{wang2024disco}. These metrics evaluate the generations at either image-level (single-level) or video-level. Image-level metrics include L1, PSNR~\cite{hore2010image}, SSIM~\cite{wang2004image}, LPIPS~\cite{zhang2018unreasonable} and FID~\cite{heusel2017gans}, while video-level metrics include FID-VID~\cite{balaji2019conditional} and FVD~\cite{unterthiner2018towards}. Following prior research~\cite{wang2024disco}, we calculated video-level metrics for sequences of 16 consecutive frames. The dataset contains around 60,000 frames from the How2Sign dataset for training and 15,000 for testing, which correspond to about 40 and 10 minutes of video, respectively. To account for variations in appearance such as clothing, we treated the same signer across different recording sessions as distinct signer identities, resulting in a total of 13 Signer IDs.

We performed several ablation studies to evaluate the efficacy of our design choices. The first ablation study focused on the rasterization function, comparing our proposed enhanced rasterization function with the simpler baseline. The second ablation experiment focused on checking frame quality. Specifically, we reported metrics for our Pose-to-Video model under three conditions: (1) ``All frames'', where no frames were excluded from training; (2) ``Valid frames'', where frames with missing landmarks were excluded from the training set, and (3) ``Proposed'', where frames with missing landmarks, blurry frames, and frames that contain landmarks that indicate temporal inconsistencies were excluded, as detailed in the final paragraph of Section \ref{subsec:module3}. 

\begin{table*}[t]
    \caption{Evaluation results of video generation (Module 3). \bm{$\uparrow$} indicates that higher values represent better performance, while \bm{$\downarrow$} indicates that lower values represent better performance. Best results in \textbf{bold}.
    \Description[This table presents evaluation results for both image-level and video-level video generation methods, comparing a baseline method against a proposed method. The table is divided into two main sections based on the type of experiment: rasterization function and frame quality check. For all columns, the first two columns, experiment and method, specify the type of experiment (rasterization function and frame quality check) and the method used (baseline, proposed for rasterization function, all frames, valid frames, and proposed for frame quality check). The rest of columns are split into image metrics and video metrics. Image metrics include L1, PSNR, SSIM, LPIPS, and FID. Video metrics include FID-VID and FVD. Key findings are: in both experiments (resterization function and frame quality check), the proposed method consistently achieves the best performance across all metrics (bolded values), with significantly lower L1, LPIPS, FID, FID-VID, and FVD scores and higher PSNR and SSIM valus compared to the baseline and other conditions.]}
    \centering
    \footnotesize
     \resizebox{\textwidth}{!}{\begin{tabular}{ll @{\extracolsep{8pt}} ccccc  cc}
    \toprule\hline
         \multirow{2}{*}{\textbf{Experiment}} &\multirow{2}{*}{\textbf{Method}}&\multicolumn{5}{c}{\textbf{Image}}  & \multicolumn{2}{c}{\textbf{Video}}\\
         \cline{3-7} \cline{8-9}
         &&\textbf{L1 $\downarrow$}  &\textbf{PSNR $\uparrow$} &\textbf{SSIM $\uparrow$} &\textbf{LPIPS $\downarrow$} &\textbf{FID $\downarrow$} &\textbf{FID-VID $\downarrow$} &\textbf{FVD $\downarrow$} \\
         \hline
         \multirow{2}{*}{Rasterization function}& Baseline &  31.71E-05&8.069  &0.028  &1.107  &401.14  & 189.62 & 2024.42\\
          &\textbf{Proposed} &  \textbf{2.83E-05}&\textbf{23.346}  &\textbf{0.864} &\textbf{0.155}  &\textbf{56.28}  &\textbf{7.23} &\textbf{173.78}\\
          \hline
         \multirow{3}{*}{Frame quality check}&  All frames& 3.60E-05 & 19.98 &0.838  &0.192  &187.03  &27.71&691.91 \\
         & Valid frames &3.14E-05  &22.03  &0.855  &0.165  &173.56  &15.72 & 497.17 \\
         &\textbf{Proposed} &  \textbf{2.83E-05}&\textbf{23.346}  &\textbf{0.864} &\textbf{0.155}  &\textbf{56.28}  &\textbf{7.23} &\textbf{173.78}\\
    \hline \bottomrule
    \end{tabular}}
    \label{tab:pose-to-video_eval_results}
    \vspace{-0.3cm}
\end{table*}

Table \ref{tab:pose-to-video_eval_results} presents the evaluation results, demonstrating the proposed approach improves all metrics across the board. The effectiveness of the rasterization function is evident, as the baseline approach produced outputs that resembled a reconstructed skeletal pose rather than a photorealistic human version. The proposed rasterization function provides a better anatomical representation of a given pose, enabling the model to learn a more robust mapping between skeletal poses and photorealistic human images. In terms of frame quality, removing lower quality frames progressively improves the model’s performance, reinforcing the conclusion that data quality is just as important as quantity.

\section{User Evaluation with DHH Signers}\label{sec:user_study}

We conducted a user study with 30 DHH participants to further evaluate our prototype system by assessing the perceived quality of our generated signed videos, with a focus on their ASL grammatical correctness---both with and without non-manual markers---understandability, and naturalness of movement. Additionally, we gathered participants' interest in this technology and its potential use cases. All English sentences were derived from continuous sentence-level signing videos in the ASLLRP dataset. The signed videos presented to participants were either generated from the How2Sign dataset or presented as raw, unprocessed human-signed videos from the ASLLRP dataset.

\subsection{Study Design}\label{subsec:study_design}
The survey was conducted online via a web-based survey tool and consisted of two main sections. Participants provided responses through 5-point rating scales and open-ended feedback, allowing for both quantitative and qualitative insights. To minimize bias that might arise from visual aesthetics influencing translation quality evaluations, we intentionally structured the survey to first evaluate visual and motion quality, followed by translation quality. This design choice was inspired by the aesthetic-usability effect, which indicates that users often perceive visually appealing or high-quality visual designs as more functional or accurate~\cite{tractinsky2000beautiful,hoegg2010good}. We chose 5-point semantic differential scales, a survey rating scale designed to capture respondents' attitudes, approaches, and perspectives~\cite{osgood1964semantic,osgood1957measurement,semantic_diff}, to gauge DHH participants' perceptions of the quality of the generated signed videos. A detailed summary of the user study questions is provided in Appendix \ref{appendix:survey}.

\subsubsection{Section 1: Visual and Motion Quality}\label{subsubsec:sec1} This section evaluated Modules 2 (ASL Representations to Skeletal Pose Sequence) and 3 (Skeletal Poses to Video Frames) of our system, focusing on the motion and visual quality of the generated signed videos. The goal was to explore alignments between technical and human evaluations while providing additional assessment of Module 2, which was not fully evaluated during the technical phase due to the lack of established metrics for this module. To achieve this, we presented two types of models for evaluation. 

The first type, \textit{AI (Annotations)}, uses human-annotated English-based glosses (manual markers) from the ASLLRP dataset, along with our manually annotated linguistic information (non-manual markers), as input for Modules 2 and 3 of our system. This approach assumes a high-quality text translation and focuses on evaluating the performance of our motion and image models. The second type, \textit{Video Retargeting}, takes skeletal poses extracted from ASLLRP sentence videos as input for Module 3, representing a best-case scenario between these two types. This approach assumes high quality text translation and skeletal extraction, focusing solely on assessing the performance of our visual model and identifying potential issues when retargeting data from the ASLLRP dataset to the How2Sign dataset. Notably, all models using our Module 3 were trained exclusively on the How2Sign dataset (detailed in Section \ref{subsec:module3}).

\begin{figure*}[t]
    \centering
    \subfloat[A video presented to participants.]{ \includegraphics[width=0.352\textwidth]{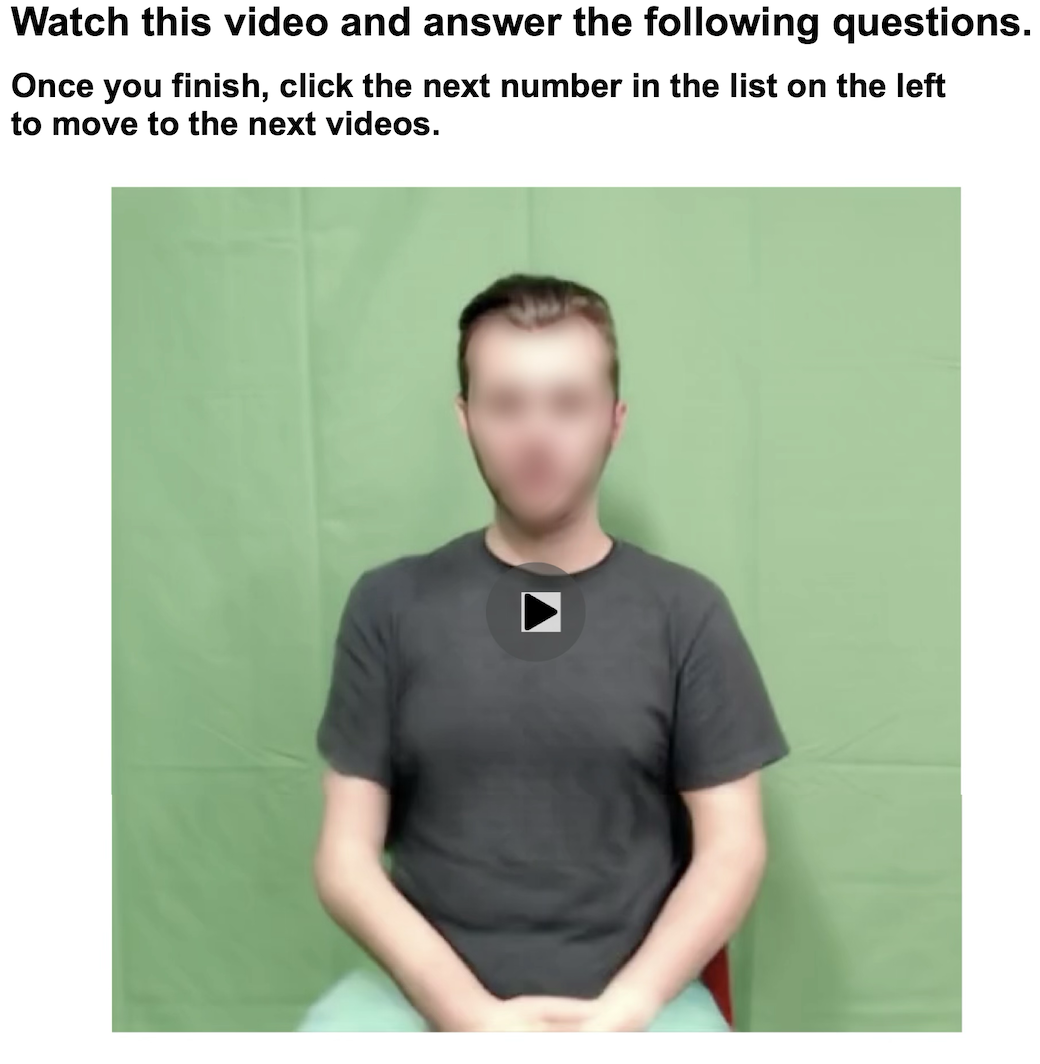}
    \label{fig:sec1_1}}
    \subfloat[Follow-up questions regarding the video.]{\includegraphics[width=0.63\textwidth]{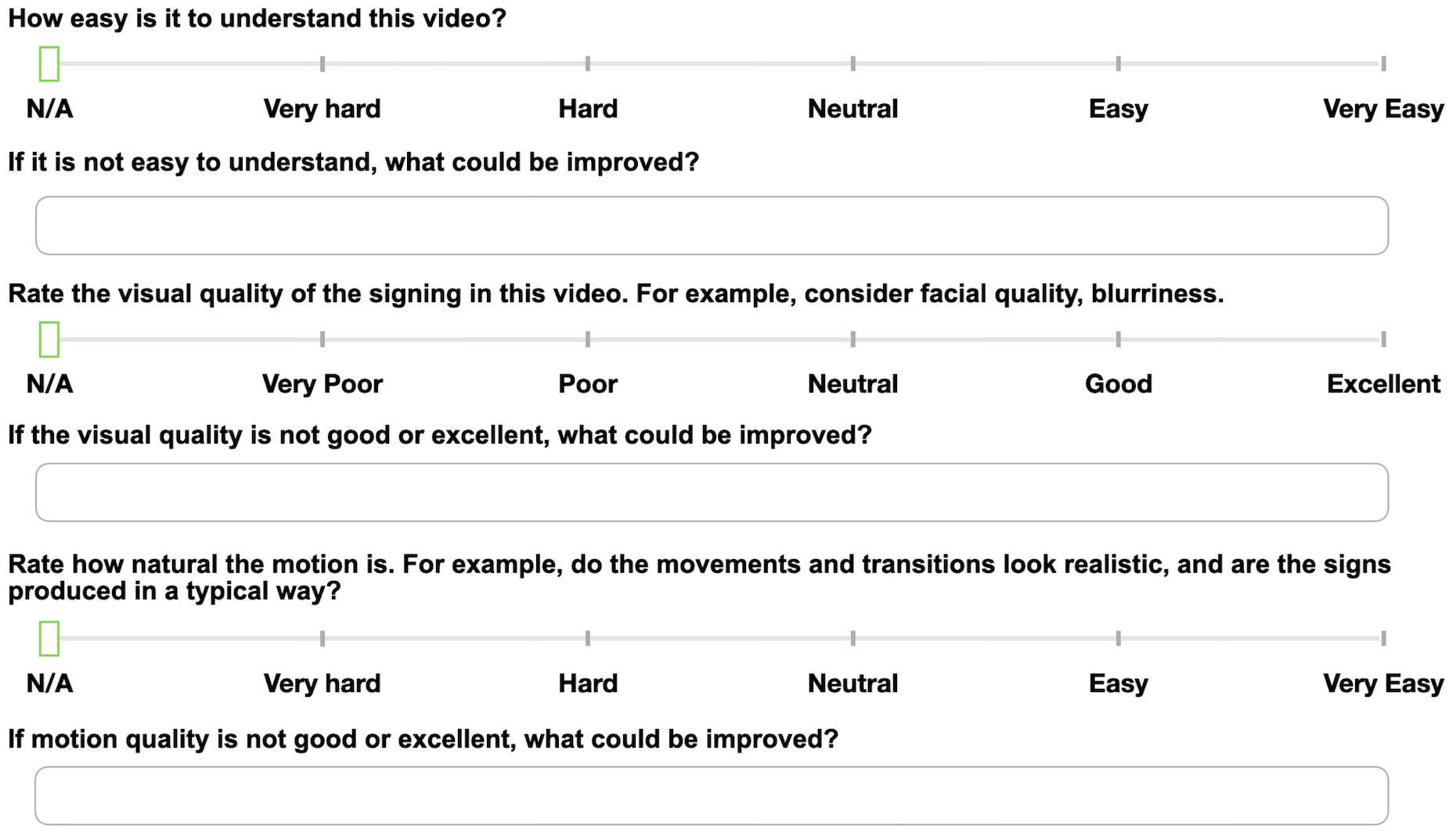}
    \label{fig:sec1_2}}
    \caption{An example screen from Section 1 of the survey. Videos generated using the How2Sign dataset were presented to participants, followed by a series of evaluation questions. The signer’s face is blurred here to preserve privacy for publication. However, participants viewed an unblurred version during the survey. 
    \Description[This figure contains an example screen from section 1 of our survey. It includes two main sub-figures. The left sub-figure is the video presented to participants, it contains a video frame with a blurred face of a generated human-realistic AI signer. The video was generated using data from the How2Sign dataset. The right sub-figure is the follow-up questions related to the video shown to participants. The questions ask participants to evaluate the video based on: how easy it is to understand (rated from very hard to very easy); the visual quality of the signing (rated from very poor to excellent); and the naturalness of motion (rated from very hard to very easy). Each question includes a scale and an optional text box for participants to suggest improvements if they rated the video negatively in any aspect.]}
    \label{fig:user_study_sec1}
\end{figure*}

Each participant viewed and rated three videos for each type, where videos were randomly sampled from a larger set of 27 sentences. Participants rated each of the videos on a 5-point rating scale for understandability, visual quality, and naturalness of movement---criteria commonly referenced in the literature~\cite{huenerfauth2007evaluating,quandt2022attitudes}. These evaluations were captured across multiple bipolar dimensions, with scale options such as ``0 (Very Hard), 1 (Hard), 2 (Neutral), 3 (Easy), 4 (Very Easy)'' or ``0 (Very Poor), 1 (Poor), 2 (Neutral), 3 (Good), 4 (Excellent).'' Note that ``N/A'' was provided as a default option, but participants were asked to select another response. After completing each rating scale question, participants had the option to provide open-ended feedback for additional insights. Figure \ref{fig:user_study_sec1} provides an example of the survey interface used for this section as presented to the participants. 


\subsubsection{Section 2: Translation Quality}\label{subsubsec:sec2}
This section evaluated Module 1 (English text to ASL Representations) of our system, focusing on the translation quality. We aimed to assess how closely our generated ASL aligns with correct ASL grammar and style, and to examine the impact of non-manual markers, specifically facial expressions, on the overall quality and comprehensibility of the signed output. We achieved this by comparing four types of models.

The first type, \textit{AI (Annotations)}, is identical to the approach described in Section \ref{subsubsec:sec1}. This approach allows us to compare the translation quality of our system with human-annotated ASL representations.  The second type, \textit{AI w/o Expr}, uses our full system (Modules 1-3) but with expression blending model turned off to specifically evaluate our system's effectiveness in generating non-manual markers. The third type, \textit{AI (Full)}, uses our full prototype with the LLM predictions (Modules 1-3), containing both manual and non-manual information. The last type, \textit{Raw Video}, consists of original human-signed videos from the ASLLRP dataset without any processing or modeling. The raw videos serve as the best-performing benchmark, providing a reference point for understanding the gap between our system-generated outputs and natural, human-signed videos.

Each participant viewed 21 videos taken from six sentence types. These included a wh-question, a yes/no question, a question that could be mistaken for a statement without non-manual markers, a simpler statement without non-manual markers, a more complex statement involving negation or conditional, and one random sentence with fingerspelling. Within a survey, videos were randomized so that the same sentence was only used once, with one exception. To analyze the expression blending part of our model, we showed each participant the three ``question'' sentences twice: once using our full system (\textit{AI (Full)}) and once without expression blending (\textit{AI w/o Expr}). 

For each video, participants first provided English translations for the ASL content shown. They were then presented with the ``true'' English translation from the ASLLRP dataset and rated three aspects on a 5-point rating scale: the similarity between the video's meaning and the ``true'' English, the quality of the ASL translation (including grammar and signing style), and the accuracy of the facial expressions. To encourage decisive responses and minimize central tendency bias, we adapted scales from prior work~\cite{zhu_neural_2023}, excluding the neutral option and using choices such as ``0 (Very Poor), 1 (Poor), 2 (Acceptable), 3 (Good), 4 (Excellent).'' After completing these ratings, participants used checkbox options and an open-ended text box to report issues with the translations. They also had the options to provide feedback on translation quality and share their ASL interpretation of the English sentence. To maintain consistency and reliability in the evaluation process, each video in both sections was reviewed by at least three participants. 


\subsubsection{Follow-Up Questions and Demographics}\label{subsubsec:sec4} At the end of the survey, participants were asked about their general interest in AI signing technology and its potential use cases. Additionally, demographic information was collected, including gender, age group, the age at which they began learning ASL, their proficiency in both English and ASL, and the frequency of their communication in ASL and spoken English.

\subsection{Data Collection}\label{subsec:data_collection}

Participants in this study were recruited outside the research group to ensure impartiality and avoid potential biases. To qualify for participation, individuals had to self-identify as DHH, use ASL as their primary language, and be over the age of 18. To ensure participants met these criteria and had the necessary proficiency in ASL, we further implemented a screening process. This process involved prospective participants watching three ASL videos and selecting the corresponding English translations from a set of multiple-choice options. This study went through our organization's internal user study review process.

After the screening and recruitment process, we enrolled a total of 30 DHH signers who met all eligibility criteria. For demographics, 11 participants were aged 20-29, 10 between 30-39, 7 between 40-49, and 2 between 50-59. Twenty-one participants identified as female, and 9 identified as male. Twenty-four participants learned ASL before age 10, while the remaining learned it later. Regarding proficiency, 23 participants rated their ASL comprehension and production as excellent, while the others rated themselves as good. Fifteen rated their English proficiency as excellent, 11 as good, and 3 as acceptable. All except one reported using ASL daily, with one reporting weekly use. The survey took 45-60 minutes for most individuals to complete.

\begin{figure*}[t]
    \centering
    \subfloat[Motion and Visual Quality.]{ \includegraphics[width=0.48\textwidth]{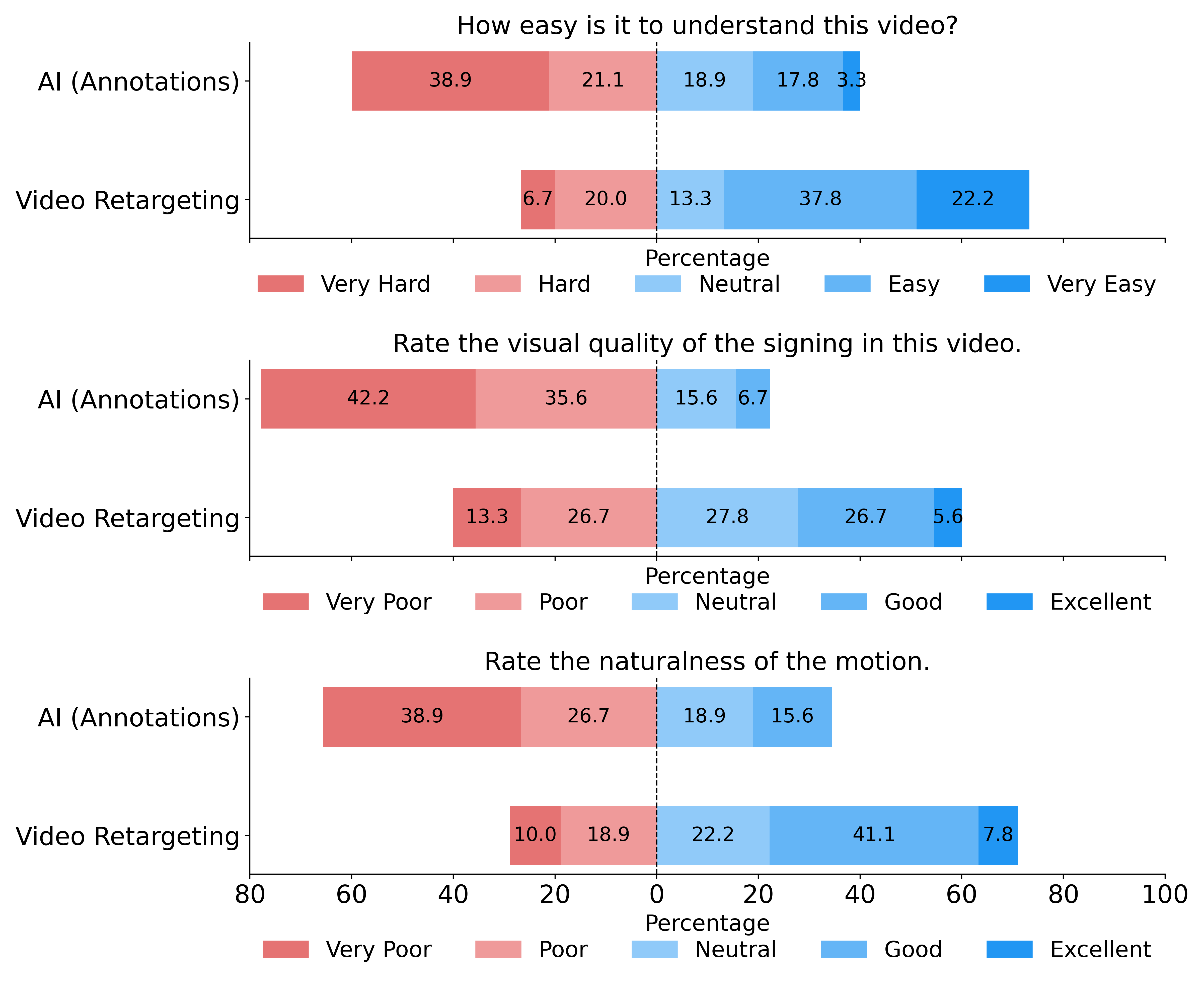}
    \label{fig:visual_motion_quality}}
    \subfloat[Translation Quality.]{\includegraphics[width=0.48\textwidth]{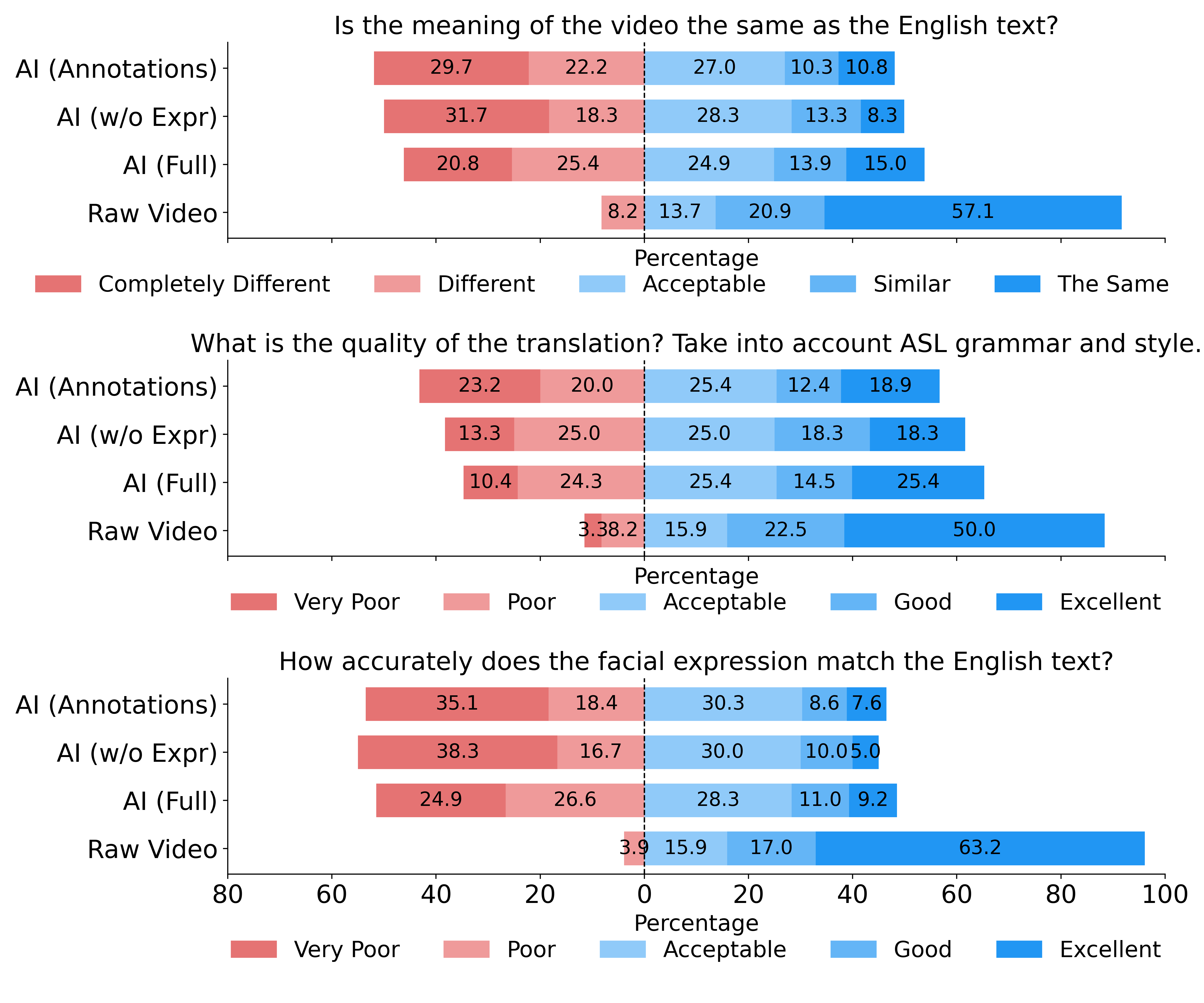}
    \label{fig:translation_quality}}
    \caption{
    Descriptive statistics summarize participants' ratings of motion, visual, and translation quality across model types.  Each bar represents the percentage of videos rated within a given response. The right side of each chart (blue) indicates a positive (or neutral) result and the left side (red) indicates a negative result. All models except \textit{Raw video} were trained on the How2Sign dataset to generate signed videos, using English sentences from the ASLLRP dataset as input. In contrast, \textit{Raw video} refers to unprocessed, human-signed videos directly sourced from the ASLLRP dataset.
    \Description[This figure presents the results from two sections of our user study evaluating video quality. It contains two sub-figures. Each sub-figure displays the percentage of responses, with the right side (blue) representing better ratings and the left side (red) representing worse rating. The left sub-figure focused on visual and motion quality and the right sub-figure focused on translation quality. In the left sub-figure, it contains three questions: how easy is it to understand this video? Rate the visual quality of the signing in this video. Rate the naturalness of the motion. In the right sub-figure, it also contains three questions: is the meaning of the video the same as the English text? What is the quality of the translation? Take into account ASL grammar and style. How accurately does the facial expression match the English text?]}
    \label{fig:userstudyresults}
\end{figure*}

\subsection{Data Analysis}

For the rating questions, we report descriptive statistics showing the proportions of each response option for each model type. To account for both fixed and random effects in our data, and to address small sample sizes and deviations from normality in data distributions, we conducted parametric bootstrap linear mixed model (LMM) analyses~\cite{davison1997bootstrap,pinheiro2006mixed}. These models include model type, sentence type, and participants' demographic variables---including gender, age category, ASL age, ASL proficiency, and frequency of ASL use---as fixed effects to assess their influence on the ratings. Participant ID was treated as a random effect to capture individual variability. For visual and motion quality evaluations, we conducted three LMM analyses--one each for understanding, visual quality, and naturalness of motion. Similarly, for translation quality evaluations, we conducted another three LMM analyses to assess the similarity of meaning between the generated videos and the English text, the signing quality (focusing on ASL grammar and style), and the accuracy of facial expressions in matching the English text. For open questions, we summarize participants' feedback to provide insight into their experiences and perceptions.

To further evaluate our system's translation quality, three authors with ASL experience (1 fluent Deaf signer; 1 fluent hearing signer; 1 novice hearing signer) independently rated the participant-provided translations relative to the English annotations from the ASLLRP dataset. This evaluation assessed whether each translation was semantically equivalent to the target phrase. A 5-point scale was used, defined as follows: 4 = the idea is the same (The same); 3 = the idea is evident but contains one error, such as question changed to a statement, one word error, or one missing element (Similar); 2 = the idea is somewhat similar but unclear or contains multiple errors (Acceptable); 1 = some semblance of the idea is present (Poor); 0 = little to no resemblance to the target (Completely different). Pairwise Pearson correlations~\cite{lee1988thirteen} were conducted and showed the high agreement among the ratings of the three evaluators, with Pearson's correlation coefficients ranging from $r = 0.860$ to $0.946$ ($p < .001$). For all LMM analyses with model type as a fixed effect, additional pairwise post-hoc comparisons with Holm corrections~\cite{holm1979simple} were conducted to identify specific factors influencing translation quality.


\subsection{Findings}\label{subsec:study_findings}

\subsubsection{Visual and Motion Quality Evaluation Findings}
Figure \ref{fig:visual_motion_quality} shows results for Section \ref{subsubsec:sec1}. Regarding the understandability of the generated signed videos from two model types, in the best-case scenario, where raw ASLLRP skeleton data was retargeted using the pose-to-video model from Module 3 (\textit{Video Retargeting)}, participants found 60.0\% of videos to be \textit{easy} or \textit{very easy} to understand, with 73.3\% to be at least \textit{neutral}. Results for \textit{naturalness} were very similar. For visual quality, perceptions were lower, with 32.3\% ratings being at least \textit{good} and 60.1\% with at least \textit{neutral}. When using our full model with human annotations from the ASLLRP dataset combined with linguistic information from our LLM (\textit{AI (Annotations)}), only 21.1\% of ratings indicated the videos were \textit{easy} or \textit{very easy} to understand. Naturalness and visual quality were both rated with lower scores compared to the \textit{Video Retargeting} approach. However, in open-ended responses, some participants commented positively about the body and face movements (\eg \textit{``Good Body Movements and some lip syncing their words (helpful for those who don't understand [the ASL sign])''}). Negative sentiments focused on issues like blurriness, cut off fingers, and the need for improved facial expressions. For example, \textit{``Blurred background is hard to read the signer''} and \textit{``[...] fingers cut off sometimes, needs more movement in the facial.''}

Our parametric bootstrap LMM analyses revealed significant main effects of model type and sentence type on understandability, visual quality, and naturalness of motion, with few demographic variables showing significant effects. For example, in visual quality ratings, model type showed a significant effect, $\chi^2(1)=54.53, p<0.001$, with a bootstrap $p$-value of 0.002, indicating that the retargeted model received significantly higher visual quality ratings than the \textit{AI (Annotations)} model. Sentence type also had a significant impact on visual quality ratings, $\chi^2(4) = 18.59, p < .001$. 
\begin{figure}[t]
    \centering
    {\includegraphics[scale=0.27]{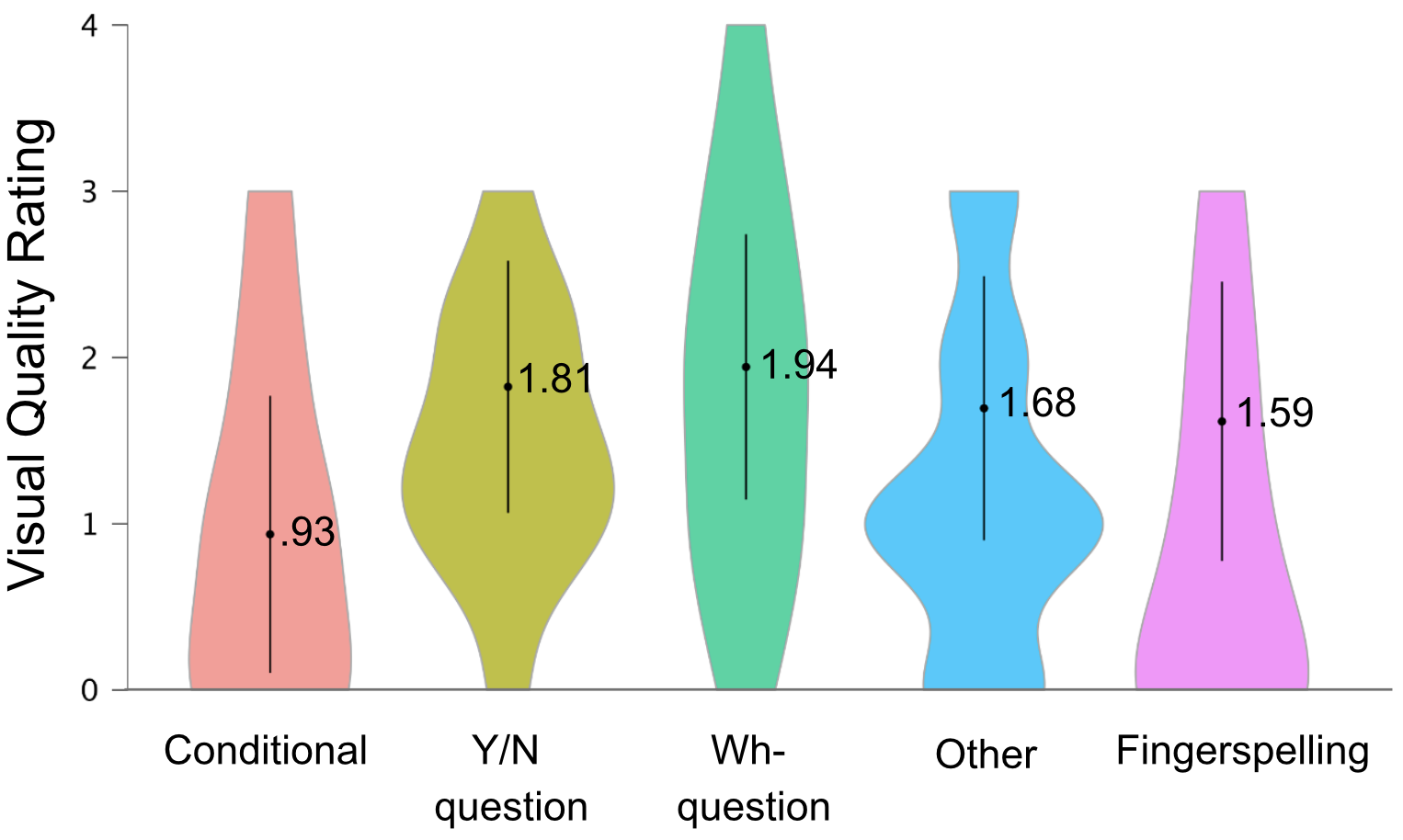}}
    \caption{Violin plot illustrating the estimated marginal means for visual quality ratings by sentence type. Wh- and yes-no questions exhibit the highest visual quality ratings, whereas conditional sentences display the lowest ratings (all p values < .001). Error bars show 95\% confidence intervals. 
    \Description[This figure presents a violin plot showing the relationship between sentence type and visual quality ratings. The x-axis categorizes different sentence types, including conditional sentences ('condition_if'), yes-no questions ('ynq'), wh-questions ('whq'), other sentence types ('other'), and fingerspelling ('fs'). The y-axis represents estimated marginal means of visual quality ratings, with higher values indicating better quality. The plot reveals that wh-questions and yes-no questions achieve the highest visual quality ratings, as indicated by their higher central markers and broader density at the top. In contrast, conditional sentences tend to have the lowest visual quality ratings, as seen in their lower central markers and density concentrated near the bottom. ]}
    \label{fig:visual_quality_sentence_type}
\end{figure}
As illustrated in Figure \ref{fig:visual_quality_sentence_type}, Wh-questions and yes-no questions were rated highest in visual quality, while conditional sentences received the lowest ratings (Holm-corrected post-hoc tests: all $z > 3.7$, all $p$ < .001). Among demographic variables, no significant effects were found for gender ($\chi^2(1) = 0.029$, p = 0.972), age ($\chi^2(3) = 1.94, p = 0.584$), ASL age ($\chi^2(2) = 2.54, p = 0.281$), or ASL proficiency ($\chi^2(2) = 2.95, p = 0.229$). 

\subsubsection{Translation Quality Evaluation Findings} Figure \ref{fig:translation_quality} shows results for Section \ref{subsubsec:sec2}. As expected, the raw videos from the ASLLRP dataset were easiest to understand and had the highest similarity with the English sentences that were shown. Surprisingly, there were a small number of ASLLRP videos that had ``poor'' or ``different'' ratings. One participant noted that one of these raw videos had a \textit{``Lack of grammar and sentence structure but I can understand what he mean[s].''} For the other three models, the differences in translation quality were modest overall---except that the results using our translated glosses (\textit{AI (Full)}) achieved significantly higher ratings than the manually annotated glosses from the ASLLRP dataset (\textit{AI (Annotations)})s in terms of the meaning of the translation. Furthermore, incorporating non-manual markers (\ie facial expressions) in our full model resulted in higher acceptance compared to the same model without non-manual markers (\textit{AI (w/o Expr)}). The \textit{quality} of the translation, which focuses on ASL grammar and style, was rated as at least \textit{acceptable} in 65.3\% of cases with our full model. Similarly, the \textit{meaning} of the translation was at least acceptable 53.8\% of the time and facial quality was at least acceptable 48.5\% of the time.


Similar to the evaluation of visual and motion quality, our LMM analyses revealed a significant main effect of model type on all three aspects of translation quality (all $p < .001$, with bootstrap $p$-values of 0.002). Contrast analyses showed that the videos generated by the \textit{AI (Annotations)} model were significantly less similar in meaning to the provided English text compared to those produced by our full model (\textit{AI (Full)}; $z = 2.73$, $p < .05$). However, raw videos consistently received higher ratings than all model-generated outputs. For signing quality and the accuracy of facial expressions, contrast analyses indicated a significant difference between raw videos and all model outputs; however, no significant differences were observed among \textit{AI (Annotations)}, \textit{AI (Full)}, and \textit{AI (w/o Expr)}. Sentence type was also identified as a significant factor influencing translation quality. For example, signing quality ratings exhibited a significant main effect of sentence type ($\chi^2(3) = 227.27$, $p < .001$, with a bootstrap $p$-value of 0.002). However, the limited sample size for each sentence type restricted the scope for more detailed analyses. No demographic variables were  significant predictors of signing quality ratings.

\begin{figure}[t]
    \centering
    {\includegraphics[scale=0.29]{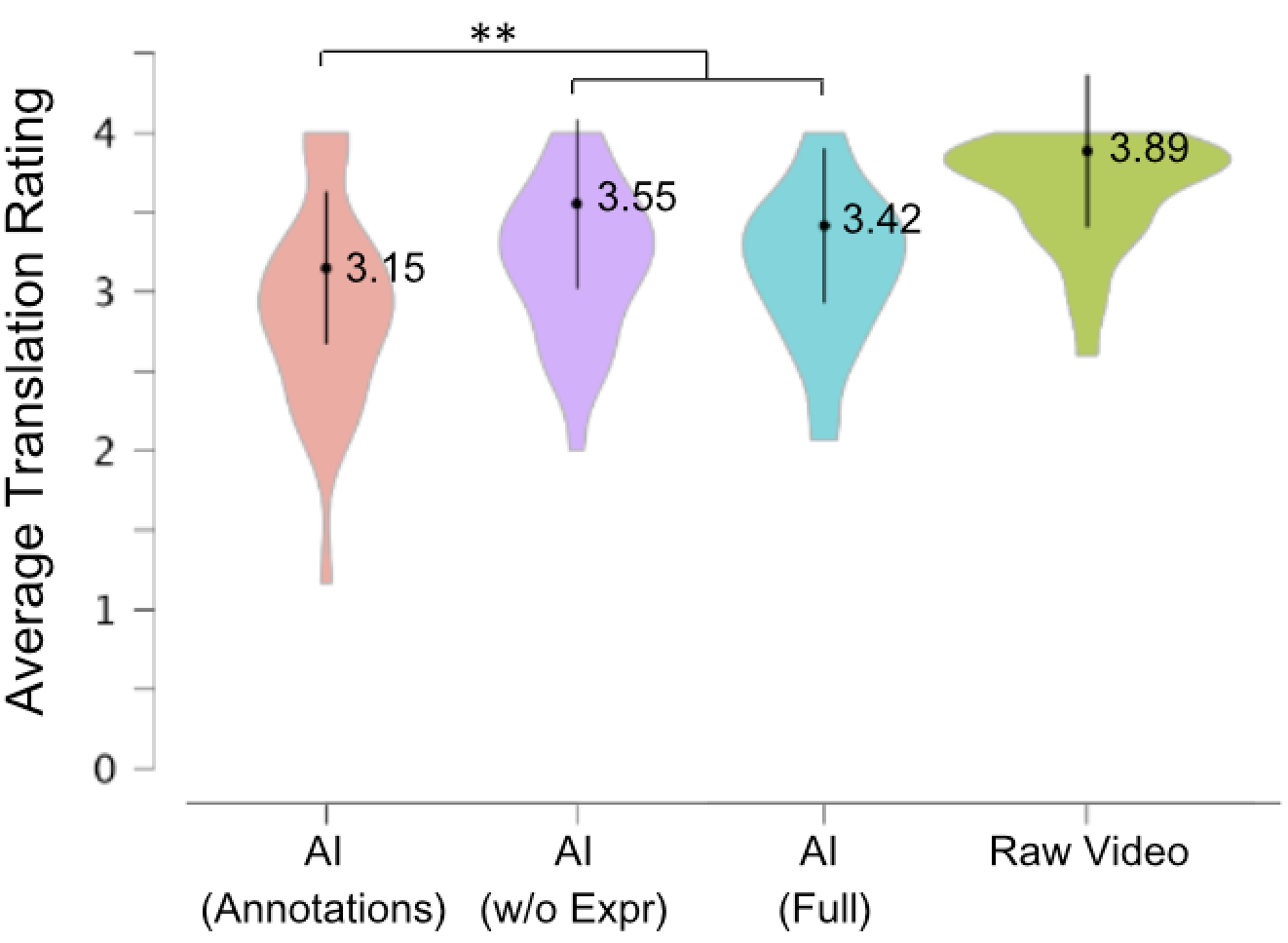}}
    \caption{Violin plot illustrating the estimated marginal means for translation quality ratings by model type. \textit{AI (Full)} and \textit{AI (w/o Expressions)} were rated significantly more accurate than \textit{AI (Annotations)}, with no significant difference between them. Error bars represent 95\% confidence intervals. While \textit{Raw Video} was rated significantly better than other models, only significance among the other three models is marked for visual simplicity.
    \Description[This figure is a violin plot displaying the estimated marginal means of Average Translation Rating for different translation methods: AI (Annotations), AI (w/o Expr), AI (Full), and Raw Video. The y-axis ranges from 0 to 4, representing the average rating. The mean rating for each method is indicated by a dot and its numerical value is displayed (e.g., AI (Annotations): 3.15, AI (w/o Expr): 3.55, AI (Full): 3.42, Raw Video: 3.89). Vertical lines show the confidence intervals around the mean. A statistically significant difference is marked between the AI (Annotations) and AI (w/o Expr) conditions, as indicated by asterisks.]}

    \label{fig:translation_quality_ratings}
\end{figure}
Our additional LMM analysis, aimed at understanding how well participants' translations aligned with the intended sentences, revealed a strong effect of model type on average translation quality ratings, $\chi^2(3) = 50.45, p < .001$, with a bootstrap $p$-value of 0.002. As shown in Figure \ref{fig:translation_quality_ratings}, the quality of the translations provided by participants showed significant differences between model types, with \textit{AI (Annotations)} performing significantly worse than both \textit{AI (Full)} and \textit{AI (w/o Expr)} ($z = 3.200$, Holm-corrected $p < .01$). Although participants rated translations with non-manual markers as more acceptable, no significant differences were observed between the two models using our system, with and without non-manual markers ($z = 0.907, p = 0.364$). Sentence type also had a significant effect on the translation quality ratings, $\chi^2(7) = 23.34, p < .001$, with a bootstrap $p$-value of 0.002. In contrast, all demographic variables did not significantly influence translation quality ratings. 




Participants reported several issues with both our full model and our model using ASLLRP human gloss annotations, with the majority of concerns pertained to image and motion quality. However, there were a small number of comments on ``missing information'' and ``wrong signs.'' One participant noted, \textit{``The signing in the beginning looks very laggy, maybe avoid spelling out the words,''} referring to limitations in fingerspelling where individual letters appeared to jump between locations or were signed  slowly compared to natural signing. 

\subsubsection{Interest of AI Signing Technology and Its Use Cases} 
Participants expressed varying levels of interest in AI signing technology. One highly enthusiastic participant remarked, 
\textit{``Everything looks good so far, most of the ASL is correct, definitely on the right path. This would be a great tool and technology for those who struggle with communication in the hearing community. It's super convenient and I can't wait. Thank you for allowing me to be a part of this,''} indicating their inclination to sign up to use such a technology in the future. The least interested person highlighted that the quality of the technology is far from being useful, stating, \textit{``I am not interested seeing AI signing technology because it's too complicated to understand the ASL signer.''} Despite this, the same participant later expressed that the technology could be valuable for certain use cases. 

When asked about their interest in photorealistic, cartoon, or 3D avatars to represent AI signers, participants provided mixed feedback, but with a lean towards photorealistic styles. One participant emphasized the value of realism, stating,
\textit{``Realistic and Authentic[---]it is simpler for viewers to relate to and believe in the content when a live signer offers an honest and realistic experience. It better for training and teaching other people ASL.''} 
Stylized signers could be of interest for social media, advertisements, or children's content, but multiple people noted the importance of ensuring the stylized depiction is capable of conveying nuance of sign language: \textit{``I think more stylized appearance can do, but needs [to be] clear in image and facial expressions.''}

Participants mentioned a wide range of potential use cases for AI signing technology. Many of these examples related to simultaneous recognition and generation of ASL for real-time social interactions. Others focused on one-sided interactions, such as ASL generation of live presentations. 
\section{Discussion}\label{sec:discussion}

Our goal was to develop a prototype ASL generation system, addressing key challenges limiting real-world applicability of existing SLG systems, and to explore whether DHH signers would find this technology useful. Below, we reflect on our design process, provide key insights learned, identify areas for improvement, and discuss computational and ethical considerations in the use of our system.

\subsection{Technical Insights from the Design and Evaluation Process}

During the design process and evaluations with DHH participants, we gained valuable technical insights that informed our choices and identified areas for future improvement. One key finding was the importance of careful data handling for translation tasks. Our ablation study results, as shown in Table~\ref{tab:text-to-gloss_eval_results}, highlight the importance of data preprocessing, increasing the number of examples provided to the model, and constraining the translation within the pre-generated vocabulary to improve model translation performance in the low-resource settings. Considerable effort was dedicated to creating an annotation scheme that not only accurately represents ASL signs and sentences but also functions effectively when used with the LLM and the rest of our prototype. This points to a fundamental challenge with glossing: the diverse definitions and interpretations of ASL glosses. Standardization across datasets could mitigate this issue and improve accuracy by allowing the combination of different data resources~\cite{bragg_sign_2019}.

Our use of an LLM for generating both manual and non-manual information demonstrate potential, with the model achieving a BLEU-4 of 0.276 for translating English sentences from the ASLLRP dataset into ASL glosses. While direct comparisons---such as running our dataset on other systems or applying our system to other datasets---are challenging due to the inaccessibility of other datasets and systems, this represents highest reported score for such translation task in the literature, highlighting the effectiveness of few-shot prompting techniques in handling low-resource languages. More than half of the time, DHH participants found the meaning of the generated videos ``Acceptable'', ``Similar'' or ``The Same'' when compared to the English text. However,  in close to 50\% of the examples, they rated our translations as ``Poor'' or ``Very poor'' concerning ASL grammar and style, indicating a need for further improvement in aligning the output with native signing conventions. 

Our additional experiments on English Text-to-ASL gloss translation using Retrieval Augmented Generation (RAG)~\cite{lewis2020retrieval} demonstrated improved performance, achieving a BLEU-4 score of 0.279 $\pm$ 0.003. These results suggest potential for further enhancement in translation accuracy. Detailed descriptions of the experiments are provided in Appedix \ref{appendix:additional_text-to-gloss}. Beyond translation accuracy, our innovation on extracting non-manual markers directly from the English text using zero-shot prompting, could potentially enhance the naturalness and grammatical accuracy of the generated videos. Nonetheless, some linguistic features were misidentified due to inconsistencies between gloss annotations and English sentences (as discussed in Section \ref{subsubsec:linguistic_predictions}), suggesting the need for prompt fine-tuning or more targeted examples. 

The use of a Motion Matching approach for generating skeletal pose sequences offered both promise and challenges. By optimizing for ``economy of motion,'' this method enabled smoother transitions between signs, contributing to more fluid and natural signing overall. However, we encountered issues with fingerspelling, where unintended movements appeared between letters, disrupting the continuity of motion. This challenge was also noted in user feedback, highlighting gaps in achieving the desired naturalness in coarticulations, particularly for complex cases such as fingerspelling. The noticeable naturalness rating difference between the full model and the retargeted approach---where only in 34.5\% of the cases participants perceived the naturalness of our videos as ``Neutral'' or better, compared to 71.1\% for the retargeted version (results shown in Figure~\ref{fig:visual_motion_quality})---emphasizes the need for refining our skeletal motion generation method. 

A key factor limiting the adoption of existing SLG systems by DHH users is the low quality of the generated signing videos, which are often described as robotic or blurry ~\cite{kipp2011assessing,tran2023us,huenerfauth2009sign,quandt2022attitudes}. Our technical evaluations, as detailed in Table~\ref{tab:pose-to-video_eval_results}, demonstrate that our approach improves the visual quality of the generated videos by systematically eliminating data errors, such as missing landmarks, blurriness, and temporal inconsistencies in landmark positioning, through using only the highest-quality frames. However, we still observe a gap between these technical improvements and practical usability, as in 77.8\% of the time DHH participants found the visual quality of our signing videos to be ``Poor'' or ``Very Poor''. Additionally, participants noted that head movements did not consistently align with the camera. 
Future work could explore integrating more advanced generative models such as diffusion models~\cite{croitoru2023diffusion,yang2023diffusion} to enhance video quality.

\subsection{A Need for Larger, High-quality, and Comprehensive ASL Datasets}

Despite using the largest and highest-quality ASL datasets available, the chosen datasets still suffer from several limitations. The ASLLRP dataset is advantageous in that it contains tens of thousands of videos with comprehensive annotations (\eg glosses, English sentences, non-manual markers). However, the dataset suffers from limited visual quality due to issues such as image resolution and motion blur, which proved challenging for generating compelling image-to-image models during our initial experiments. When we turned to the How2Sign dataset for training image-to-image models, we found that the visual quality was significantly better. However, this introduced inconsistencies between datasets. For example, signers in the ASLLRP dataset tend to be seated and looking at prompts away from the camera; while signers in the How2Sign dataset maintain direct eye contact and are looking closer at the camera, slightly sideways. These discrepancies, coupled with the relatively small size of these datasets, highlight the need for more comprehensive and consistent ASL datasets.

Furthermore, the reliance on human-labeled gloss annotations in existing ASL datasets introduces multiple sources of errors and inconsistencies. 
While many English sentences in the ASLLRP dataset are derived from context-free ASL utterances translated into glosses and English text, others come from longer narrative videos. In these cases, accurate translation requires full contextual understanding, which the annotations may not always provide~\cite{tanzer2024reconsidering}. Consequently, for many of these context-dependent sentences, our text-to-gloss translations may be more accurate than the original human annotated glosses. This is reflected in our model's performance, where we achieved a BLEU-4 score of 0.305 without these context-dependent sentences (52 in the test set), compared to 0.276 with them. Further supporting this observation, our user study indicated that DHH users rated the quality of our translations, specifically regarding the meaning of video compared to the English text, to be more acceptable than the manually annotated glosses provided by the ASLLRP dataset. These findings highlight the critical need for more robust, high-quality datasets with standardized annotation practices to support the development of effective SLG systems~\cite{bragg_sign_2019}.

\subsection{Addressing the Complexities of ASL in Sign Language Generation Technologies}

The complexities of ASL grammar present challenges for developing effective SLG technologies. While general guidelines for ASL grammar exist, the language, like all natural languages, does not always adhere to rigid grammatical structures in everyday use. This complexity is evident in the mixed results from our experiments, where attempts to provide grammar guidelines to the LLM did not consistently enhance translation performance. Many examples in the ASLLRP dataset, while grammatically correct, diverge from these general guidelines (as shown in Table \ref{tab:text-to-gloss_eval_results}). Feedback from our user study, which highlighted stylistic and grammatical errors, emphasizes the need for a more nuanced computational understanding of how ASL is used in diverse, real-world contexts to improve. Furthermore, regional variations within a single language and differences across multiple sign languages introduce additional layers of complexity that remain to be addressed. 
 
This work studies several aspects of both manual and non-manual markers in ASL morphology, lexicon, and syntax, such as compounds, agreement verbs (directional verbs indicating agreement with the subject and object), fingerspelling, and name signs (more details can be found in Table~\ref{tab:gloss_convention}). However, these linguistic features are analyzed only within the context of the dataset used in this study, which does not capture the full range of their usage in ASL. Additionally, several other facets of ASL grammar and usage remain unexplored. For example, we excluded one type of manual marker, classifiers, due to limited data available to model them accurately. Classifiers, which are essential for conveying nuanced meanings and spatial relationships in ASL, require context-aware data and more sophisticated modeling approaches. As SLG systems evolve towards context-dependent applications, incorporating classifiers will be critical for enhancing the naturalness and expressiveness of the generated signs. Additionally, our work focuses primarily on eyebrow movements, one type of facial expressions within non-manual markers, used to indicate questions, conditional statements, and negation. However, non-manual markers in ASL consist of a wide range of features, including head tilts, mouth shapes, and body posture, which also contribute to the grammar and meaning of signed sentences~\cite{Stokoe1961SignLS,klima1979signs,brentari2002prosody}. Future work is needed to expand the modeling of these additional markers to capture the full complexity of ASL.

Moreover, our study focused on context-free SLG, where each sentence is generated independently. However, sign languages heavily use indexing and spatial referencing, such as referencing people or places mentioned earlier in a conversation~\cite{winston1991spatial,friedman1975space}. Our current prototype system lacks the capacity to remember or track these spatial references over multiple utterances. Additionally, types of signing like storytelling often involve more extensive use of expressions, classifiers, spatial references, and role shifting than our prototype can currently support. Addressing these challenges will require more data, modeling, and interdisciplinary collaboration with ongoing feedback from the DHH and signing communities. 

\subsection{Computational and Ethical Considerations}

While both technical and human evaluations demonstrate the potential of our prototype system, and the modular approach offers flexibility by enabling individual components to be improved or replaced as technologies advance, there are several computational and ethical considerations that should be carefully addressed when using or further improving the system. First, the current prototype requires running GPT-4o inference for every generation instance with longer prompts, which introduces computational and financial costs, as well as scalability challenges, particularly for real-time or large-scale applications. Optimization techniques or lighter models may need to be explored to address this issue. Second, the nature of modular approach can lead to the loss of information between stages, computational inefficiencies, or biases imposed by external constraints at each module. Addressing these shortcomings will require careful integration of modules. Third, the use of LLMs might pose a risk of generating inappropriate or offensive language, which could introduce harm to the DHH community or undermine their trust in using such system. As emphasized in both academia and industry (\eg Apple's Responsible AI white paper~\cite{applewhitepaper}), designing AI tools with care to proactively mitigate potential harms must be a top priority. This includes implementing content filtering mechanisms, rigorous validation processes, and culturally sensitive design practices to ensure that the system outputs are respectful, inclusive, and aligned with community expectations.



\section{Conclusion}\label{sec:conclusion}

In this paper, we proposed a prototype ASL generation system aimed at improving the naturalness, comprehensiveness, and overall quality of generated signs, addressing key limitations in existing approaches. Our technical evaluations indicate that our proposed approaches improve these aspects, enhancing the quality of generated ASL content. Feedback from DHH participants was mixed; while there was general interest in the system, concerns regarding visual quality and naturalness were noted. Reflecting on our design process and study findings, we discuss key insights and identify key areas for future improvement. While further work is needed, our study takes an initial step toward developing sign language generation systems that better meet the needs of the DHH and signing communities, offering real-world value.


\begin{acks}
  We sincerely thank all reviewers for their valuable feedback, which significantly enhanced our work. We also extend our gratitude to the participants of our user study for their time and contributions. Lastly, we deeply appreciate Gus Shitama, Julia Sohnen, Pooja Solanki, Sheridan Laine, and Antony Kennedy for their insightful discussions and support with study-related tasks. 
\end{acks}

\bibliographystyle{ACM-Reference-Format}
\bibliography{zotero_asl} 

\newpage

\appendix

\section{A Review of American Sign Language and Publicly Available Datasets}\label{appendix:ASL}

Similar to other sign languages, ASL is also a visual-based natural language, expressed by using both manual and non-manual markers~\cite{Stokoe1961SignLS}.  A common misconception is that substituting each written English word with a corresponding ASL sign would be enough as a translation~\cite{aslgrammar}. However, this approach does not produce true ASL~\cite{hanson2012computers}, as ASL has its own grammar and lexicon, distinct from English~\cite{lucas2001sociolinguistics,valli2000linguistics}. Moreover, there is no one-to-one mapping between English words and ASL signs, which makes direct substitution less appropriate~\cite{neidle2007signstream}. 


\subsubsection{ASL Written Representation}
ASL-LEX~\cite{caselli2017asl} has been used as a gloss reference for annotation of ASL in several works (\eg~\cite{desai2024asl,joze2018ms,ma2018signfi,bragg2021asl}). However, ASL-LEX glosses often lack representation of non-manual markers, such as facial expressions and body movement, which can limit the naturalness and understandability of generated signs when used in SLG~\cite{huenerfauth_evaluation_2008}. To address this, ASL linguists have developed conventions to capture non-manual markers in addition to manual behaviors~\cite{neidle2001signstream,neidle2007signstream,neidle2002signstream}. These include behaviors such as head position and movements, eye gaze and aperture, eyebrow position and movements, and body movements. 


\subsubsection{ASL Datasets} 
\label{subsubsec:asl_datasets}
Sign language datasets often pose a bottleneck for SLG research~\cite{bragg_sign_2019}. Reviewing ASL datasets reveals substantial variation in vocabulary size, recording duration, number of signers, image resolution, modalities, gloss annotation conventions, and annotation tools~\cite{zahedi2005combination,dreuw2008benchmark,martinez2002purdue,joze2018ms,neidle2012new,neidle2022asl,desai2024asl,signingsavvy,duarte_how2sign_2021,shi2022open,uthus2023youtubeasl} (Table \ref{tab:asl_datasets}). For instance, OpenASL~\cite{shi2022open} and YouTube-ASL~\cite{uthus2023youtubeasl} stand out with their extensive vocabularies of approximately 33,000 and 60,000 signs, respectively, offering a broad lexical base. However, these datasets provide only videos and English captions, without their corresponding written representations. 

RWTH-BOSTON-50~\cite{zahedi2005combination} and Purdue RVL-SLLL~\cite{martinez2002purdue} are among the earliest publicly available ASL datasets. Despite their pioneering role, their relatively small vocabularies, lack of detailed gloss annotations, non-expert human annotators, and variable image quality limit their utility for more advanced ASL research and applications. MS-ASL~\cite{joze2018ms} and ASL Citizen~\cite{desai2024asl} provide word-level isolated ASL signs from a wide range of signers, serving as valuable resources for sign language recognition research. However, for tasks such as generating ASL signs from English sentences, word-level datasets lack crucial contextual information, such as sentence structure, non-manual markers, and signer consistency.

Datasets like NCSLGR~\cite{neidle2012new}, ASLLRP\cite{neidleboston}, and DSP~\cite{neidle2022asl}, resulting from collaborations among multiple universities, as well as the How2Sign~\cite{duarte_how2sign_2021} dataset collected with higher resolution cameras, offer more comprehensive data. These datasets include English sentences with corresponding written representations, detailed annotation conventions (\eg \cite{neidle2001signstream,neidle2007signstream}), and videos featuring both continuous and citation-form signs. These advancements have allowed some of these datasets, such as NCSLGR and How2Sign datasets, to be used as benchmarks for ASL processing research (\eg~\cite{zhu_neural_2023,moryossef_data_2021,baltatzis2024neural}). While these datasets address some of the critical gaps in earlier resources, issues such as their relatively small sizes (\eg \cite{neidle2012new,neidle2022asl}), inconsistent annotation conventions across datasets, and limited accessibility of the DSP and How2Sign gloss datasets make some tasks of ASL processing both promising and challenging. 

\section{Module 1: English Text-to-ASL Gloss}
\subsection{Data Preprocessing}\label{appendix:data_prep}
\paragraph{Step 1: Data Extraction} We obtained the ASLLRP dataset from the project web interface\footnote{DAI 2: \url{https://dai.cs.rutgers.edu/dai/s/cart}, login required.}. The dataset includes ASL sentence-level signed videos and XML files\footnote{These XML files are generated from the SignStream annotation tool. More details about these files can be found here: \url{https://dai.cs.rutgers.edu/downloads/XML-Export-format.pdf}.} containing corresponding English translations and annotations. For the translation task in Module 1, we focused on extracting manual information from the textual annotations to capture the primary meaning of the English translations. Specifically, we extracted existing English sentences from the XML files and systematically spliced English-based annotations, including vocabulary and compound symbols, fingerspelling, name signs, classifiers, locative words, and gestures, in chronological order. In total, we extracted 2,119 English sentences with corresponding English-based glosses. Additionally, we trimmed the signing videos based on the XML data so that each English sentence corresponds to a specific sign language video (utterance) for our subsequent tasks.

\paragraph{Step 2: Data Cleaning} Following a similar approach to prior work~\cite{amin_sign_2021}, we removed gloss annotations that did not alter the overall meaning of the sentences when omitted, such as repetition (annotated as a single or multiple ``+'' signs), number of signing hands (annotated as ``(1h)'' and ``(2h)''), and signs indicator that both hands move in an alternating manner (annotated as ``alt.''). To reduce translation errors, we standardized all fingerspelling-related glosses from fs-XXX to fs-X-X-X (\eg from ``fs-JOHN'' to ``fs-J-O-H-N'') and unified annotations for spatial locations (\eg ``i:GIVE:j'' and ``i:GIVE:k'' were standardized to ``i:GIVE:j''). While classifiers play a crucial role in ASL, we excluded them from this work because they typically appear only once or very few times in the datasets, so there was insufficient data for effective model prompting. After data cleaning, we retained 1,843 English sentences with corresponding English-based glosses for the remaining experiments. 

\paragraph{Step 3: Text-to-Gloss Dictionary Generation} To improve consistency in sign representations across different sentences and datasets, we constructed a text-to-gloss dictionary using the ASLLRP Sign Bank\footnote{\url{https://dai.cs.rutgers.edu/dai/s/signbank}}, which contains isolated signs along with their corresponding English-based glosses and translations. We then systematically unified the glosses based on step 2 to ensure consistency between the dictionary and the gloss annotations for the sentences. During the dictionary generation, we observed that some words may have variants of glosses depending on the context (\eg ``ask, inquire, query, question'' can be annotated as ``ASK'', ``ASK:i'', or ``i:ASK:j'', depending on whether the previous and following words are signed in a neutral location). Therefore, our dictionary employs a one-to-multiple mapping, accommodating the variability in gloss annotations. In total, the dictionary contains 3,915 text-to-gloss pairs. Notably, we identified 43 words that do not have corresponding glosses (\ie out-of-vocab words). For these words, which lack corresponding videos, fingerspelling is used as an alternative. 

\paragraph{Step 4: Ground True Correction} During the process of extracting ground truth from XML files to determine whether a sentence is a yes/no question, wh- question, conditional statement, and/or contains negation, we discovered that the ground truth labels were based on the signing rather than the English text, leading to some misalignments between the English text and the linguistic labels. For example, ``I guarantee that the parents will be mad if the children dye their hair orange'' was originally labeled as a negation statement, because the signing of it contains negation, although the English sentence does not. To address these issues, four of our researchers iteratively re-labeled and discussed the test set sentence categories, refining the labels to better reflect the text content. These revised labels were then used as the ground truth, allowing us to calculate precision and recall for each sentence type predictions and to identify patterns in the model's errors.

\subsection{ASL Grammar Guidelines for LLM Prompt}\label{asl_grammar_rules}

\lstset{
  basicstyle=\small\normalfont\sffamily,    
  numbersep=5pt,                         
  tabsize=2,                              
  breaklines=true,                        
  stringstyle=\ttfamily,
  showspaces=false,
  showtabs=false,  
  linewidth=\linewidth,
 }

\begin{lstlisting}[language={},breakindent=10pt]
American Sign Language (ASL) commonly uses a type of sentence structure called topicalization. Topicalization is when the topic of a sentence is placed at the beginning of the sentence. For instance, in English, the topicalized form of the sentence, "I see my friend" would be "My friend, I see them". This is often referred to in ASL as topic/comment structure. Any description of the topic, such as including adjectives, would also come before the comment. The sentence "I see a big orange cat" would be signed as follows: CAT ORANGE BIG IX-1p SEE.

As a very visual language, ASL often requires signers to visualize a sentence and arrange their signs accordingly. Sentences that involve cause-and-effect statements, real-time sequencing, or general-to-specific details follow a specific pattern. Cause-and-effect sentences in ASL tend to place the cause before the effect in the sentence. For example, in the statement "I feel calm when I go to the park", the cause of "go to the park" would be expressed before the effect of "I feel calm". The sentence would be signed as: PARK GO-TO FEEL CALM ME.

Some sentences involve real-time sequencing, where events must be arranged in chronological
order according to how they happened in real time. For instance, the sentence "I'm worried
because my brother didn't call me after he left" would be rearranged as: POSS-1p BROTHER LEAVE CALL-BY-PHONE-1p NOT CONCERN IX-1p.

In sentences where a signer is setting a scene, the signer should move from general to specific
details. For example, in the statement "I am excited after moving to my new house in Virginia",
the signer would begin with the biggest detail ("Virginia") and work their way down to the
smallest detail ("I"). The sentence would be signed as: VIRGINIA HOUSE NEW MOVE FINISH EXCITED IX-1p.

Verbs are not conjugated based on tense in ASL, so every verb is in its base form. This means that "ate", "eats", "eating", and "eaten" are all expressed by the sign EAT. The tense is established separately by including a time indicator in the sentence. Time signs are usually placed at the beginning of the sentence, before the topic, which tells the watcher when the rest of the sentence takes place. Signers can also express tense using a sign that relates the progress of the activity, like in the image above, which uses the FINISH sign to indicate that the action is in the past and translates to "I saw".

Basic sentence structure in ASL follows the pattern of Time + Topic + Comment. The word order can change depending on the needs of the signer, but this is the most common format.
    - Time = Any necessary time indicators (establishes tense)
    - Topic = The main focus of the sentence (a noun)
    - Comment = What is being said about the topic (includes the verb)
For example,in English, one might say, "I went to the library yesterday." In ASL, the sentence might be structured like this: 
    - Time = YESTERDAY
    - Topic = LIBRARY
    - Comment = IX-1p GO-TO
As is the case with English sentence structure, sign choice and order often vary based on
context. The example above is shown in Object-Subject-Verb (OSV) order, in which the object
(the library) is the topic. However, the sentence can also be arranged in Subject-Verb-Object
(SVO) order, in which "I" is the topic and "GO-TO LIBRARY" becomes the comment: YESTERDAY IX-1p GO-TO LIBRARY

Both sentences are grammatically correct, and different factors can influence which structure the signer chooses, such as how familiar the watcher is with the library, and therefore what level of emphasis is needed.

When a question is asked in ASL, the WHO, WHAT, WHEN, WHERE, WHY, WHICH, or HOW sign is located at the end of the sentence, or if emphasis is needed, both the beginning and the end. This word order reflects topic/comment structure. For example, in English, one might ask, "What is your name?" In ASL, the sentence would be structured in this way: YOUR NAME WHAT

Additionally, while English often employs different forms of the verb "to be" in sentences, this
verb is not used in ASL and should not be included in signed conversations. 

When using negating signs in a sentence, such as NOT or NONE, the negative sign typically follows the word it is negating. For example, "I don't have any pets" would be signed as: PET HAVE NOT.
\end{lstlisting}

\subsection{Experiments on English Text-to-ASL Gloss}\label{appendix:llm_experiments}
\subsubsection{Model Selection} We experimented with various versions of GPT and tested multiple configurations to identify the optimal model. As shown in Table \ref{tab:llm_experiment_results}, GPT-4o-2024-05-13 (our adopted model) outperformed other GPT-4 variants under identical settings. Additionally, we fine-tuned two versions of GPT models capable of fine-tuning, but their performance was lower than that of few-shot prompting with the adopted model. However, fine-tuning GPT-4 models with larger datasets could hold promise, and exploring this option when the feature becomes more widely available may yield further improvements.

\subsubsection{Prompting Examples} For Module 1, we varied the prompts for the ``SYSTEM'' in different setups for the English Text-to-ASL Gloss task (depicted on the left side of Figure \ref{fig:module_1}), while maintaining consistency in the ``ASSISTANT'' and ``USER'' prompts. No additional prompt engineering was performed for generating linguistic information (task on the right side of Figure \ref{fig:module_1}). A summary of these setups is provided in Table \ref{tab:prompt_engineering}.

\subsection{Additional Experiments on English-to-Gloss Translation}\label{appendix:additional_text-to-gloss}
To enhance our translation capabilities, we implemented Retrieval Augmented Generation (RAG)~\cite{lewis2020retrieval} with anonymized embeddings. First, as a pre-process, we anonymized all train sentences by converting name references into pronouns. Next, we embedded the anonymized sentences using an OpenAI embeddings model. Finally, at inference, for each test sentence, we embedded it as well and look for the $N$ most similar examples to this sentence based on the cosine similarity between the embedding of the test example, and the embeddings of the anonymized train examples. This way, the model is presented with the most accurate and relevant examples. As Table~\ref{tab:text-to-gloss_RAG_eval_results_appx} shows, when using RAG the results are better than using all of the train examples. Moreover, using fewer examples and anonymized embeddings yields better results in most cases. The reason for using anonymization, is that names are given high weight in the embedding, which leads to less relevant examples in some cases. For examples, the 3 most similar sentence for the sentence "Which college did Mary go to?" before anonymization, are: "Which college does Mary go to?", "What did Mary's name used to be?", "Mary used to live in Boston.", While after anonymization they are: "Which college does Mary go to?", "Which high school did you go to?", "Where did you go to high school?", which are more relevant and similar examples.

\subsection{Summary of Existing Results}\label{appendix:existing_text-to-gloss_results}

Unlike German datasets such as RWTH-PHOENIX-Weather 2014T~\cite{camgoz_neural_2018} and the public DGS corpus~\cite{hanke2020extending}, which are widely used and frequently reported in the literature~\cite{chen2022two,yin2023gloss,li2020tspnet,saunders_progressive_2020}, there is comparatively less work utilizing ASL datasets. We summarize the existing translation results for ASL in Table~\ref{tab:text-to-gloss_sota}.

\section{Survey}\label{appendix:survey}
\subsection{Section 1 (Visual and Motion Quality)}
\begin{itemize}
    \item How easy is it to understand this video? (0  = Very Hard, 1 = Hard, 2 = Neutral, 3 = Easy, 4 = Very Easy)
    \item If it is not easy to understand, what could be improved? (Open-ended question)
    \item Rate the visual quality of the signing in this video. For example, consider facial quality, blurriness. (0 = Very Poor, 1 = Poor, 2 = Neutral, 3 = Good, 4 = Excellent)
    \item If the visual quality is not good or excellent, what could be improved? (Open-ended question)
    \item Rate how natural the motion is. For example, do the movements and transitions look realistic, and are the signs produced in a typical way? (0 = Very Poor, 1 = Poor, 2 = Neutral, 3 = Good, 4 = Excellent)
    \item If motion quality is not good or excellent, what could be improved? (Open-ended question)
\end{itemize}

\subsection{Section 2 (Translation Quality)}

\begin{itemize}
    \item Translate the ASL in this video into English. (Open-ended question)
    \item The intended English sentence was: ``Do you have to work all night? (example)'' How similar is the meaning of the video compared to the English text? (Completely Different, Not Similar, Acceptable, Similar, The Same)
    \item What is the quality in the ASL translation? Take into account ASL grammar and signing style but not the visual fidelity. (0 = Very Poor, 1 = Poor, 2 = Acceptable, 3 = Good, 4 = Excellent)
    \item How accurately does the facial expression match the English text? (0 = Very Poor, 1 = Poor, 2 = Acceptable, 3 = Good, 4 = Excellent)
    \item Did any of the following make the video harder to understand? (Multi-choice)
    \begin{itemize}
        \item Grammars/sentence structure
        \item Wrong signs
        \item missing information
        \item Wrong facial expressions
        \item Lack of image clarity
        \item Poor motion quality
        \item Other (write below)
        \item There were no issues
    \end{itemize}
    \item If you choose other, what else made it hard to understand? (Open-ended question)
    \item If this video does not convey the English well, how would you interpret the English sentence into ASL? Write out glosses or describe how you would sign it in ASL. (Open-ended question)
\end{itemize}

\subsection{Follow-up Questions and Demographics}

\begin{itemize}
    \item The following questions ask about your interest in AI Signing technology. First, imaging a version of this technology that is ``nearly perfect,'' meaning the videos are understandable, natural, and accurate. Answer the following with this perfect technology in mind.
    \begin{itemize}
        \item Can you image using this technology to supplement existing live interpreters, for example they were not available or for use cases where interpreters might not be possible. (Never, Rarely, Maybe, Sometimes, Often)
        \item Where might you be interested in seeing AI Signing technology? What specific applications or use cases? (Open-ended question)
        \item Why are you interested in these use cases? (Open-ended question)
    \end{itemize}
    \item All of the videos in this study are meant to look like a live ASL signer. THere are alternatives, for example if the human was stylized or had a cartoon-like look. What is your interest in these styles? Assume that both versions would be capable of all signing motions needed for ASL. (All open-ended questions)
    \begin{itemize}
        \item For what applications or purposes, if any, would you prefer video with the ``live ASL signer'' look? 
        \item Why do you think this? 
        \item For what applications or purposes, if any, would you prefer video that looked like a cartoon or 3D avatar? 
        \item Why do you think this? 
    \end{itemize}
    \item Demographics
    \begin{itemize}
        \item What is your gender?
        \begin{itemize}
            \item Woman
            \item Man
            \item Non-binary
            \item Prefer not to disclose
            \item Prefer to self-describe: $\_\_\_\_\_\_\_\_\_\_$
            
        \end{itemize}
        \item What is your age range?
        \begin{itemize}
            \item 20 to 29
            \item 30 to 39
            \item 40 to 49
            \item 50 to 59
            \item 60 to 69
        \end{itemize}
        \item At what age did you learn ASL?
        \begin{itemize}
            \item Under age 10
            \item 11 to 20
            \item 21 to 30
            \item 31 to 40
            \item 41 to 50
        \end{itemize}
        \item What is your ASL understanding and production proficiency? (Very Poor, Poor, Acceptable, Good, Excellent)
        \item What is your English reading and writing proficiency?(Very Poor, Poor, Acceptable, Good, Excellent)
        \item How often do you communicate with ASL? (Never, Monthly, Weekly, Daily)
        \item How often do you use spoken English? (specifically, you voicing to others) (Never, Monthly, Weekly, Daily)
    \end{itemize}
\end{itemize}

\onecolumn
\begin{table*}[t]
\caption{Existing ASL datasets. SL stands for sign language. ``-'' represents relevant information was not provided. ``Unknown'' represents relevant information was not found. \label{tab:asl_datasets}}
\renewcommand{\arraystretch}{1.1}
   \resizebox{\textwidth}{!}{
\begin{tabular}{L{0.15\textwidth}|R{0.06\textwidth}|R{0.06\textwidth}|R{0.06\textwidth}| C{0.1\textwidth}| L{0.12\textwidth}|L{0.45\textwidth}|C{0.15\textwidth}}
\toprule\hline
\multicolumn{1}{c|}{{\textbf{\makecell[c]{Dataset}}}} & \multicolumn{1}{c|}{{\textbf{\makecell[c]{Vocab.}}}}  &
\multicolumn{1}{c|}{{\textbf{\makecell[c]{Hours}}}} &  \multicolumn{1}{c|}{{\textbf{\makecell[c]{Signers}}}} & \multicolumn{1}{c|}{{\textbf{\makecell[c]{Resolutions \\ (pixels)}}}} & \multicolumn{1}{c|}{{\textbf{\makecell[c]{Modalities}}}} &\multicolumn{1}{c|}{{\textbf{\makecell[c]{Gloss Labeling Standard}}}} & \multicolumn{1}{c}{{\textbf{\makecell[c]{Annotation \\ Tools}}}}  \\ \hline
RWTH-BOSTON-50~\cite{zahedi2005combination} & 50 & >9 & 3& 195 $\times$ 165 & Video, word & - & - \\\hline
Purdue RVL-SLLL~\cite{martinez2002purdue} & 104 & 14  &14 & 640 $\times$ 480 & Video, Gloss & Glosses include manual English-based labels, and non-manual behaviors such as handshapes and motions for two hands.  & Human Annotator \\\hline
RWTH-BOSTON-400~\cite{dreuw2008benchmark} & 483 & - & 5& 648 $\times$ 484 & Video, Gloss, Utterance & Glosses include manual English-based labels and non-manual behaviors, both anatomical (\eg raised eyebrows) and functional (\eg wh-questions). Glosses do not include handshape annotations. & SignStream$^@$2~\cite{neidle2001signstream} \\\hline
MS-ASL~\cite{joze2018ms} & 1K & 24 & 222  & 224 $\times$ 224 & Video, Pose, Word & Glosses were generated by referencing ASL Tutorial books~\cite{zinza2006master,caselli2017asl}. & Human Annotator \\\hline
DSP~\cite{neidle2022asl}& >1.7K & - & 15 & - & Video, Gloss, Utterance, Word & Glosses include manual English-based gloss labels, sign type, start and end handshapes (both hands), grammatical markers (\eg questions, negation, topic/focus, conditional, relative clauses), and anatomical behaviors (\eg head nods/shakes, eye aperture, gaze). & SignStream$^@$3~\cite{neidle2017user} \\\hline
NCSLGR~\cite{neidle2012new}& 1.8K & 5.3   & 4 & - & Video, Gloss, Utterance & Glosses include manual English-based labels and non-manual behaviors, both anatomical (\eg raised eyebrows) and functional (\eg wh-questions). Glosses do not include handshape annotations. & SignStream$^@$2~\cite{neidle2001signstream}\\\hline
ASLLRP~\cite{neidle2022asl} & >2.7K & 3.6   & 4 & - & Video, Gloss, Utterance, Word & Glosses include manual English-based gloss labels, sign type, start and end handshapes (both hands), grammatical markers (\eg questions, negation, topic/focus, conditional, relative clauses), and anatomical behaviors (\eg head nods/shakes, eye aperture, gaze). & SignStream$^@$3~\cite{neidle2017user}\\\hline
ASL Citizen~\cite{desai2024asl}& >2.7K & 30.5 & 52 & - & Video, Gloss, Pose, Word & Glosses include manual English-based labels by referencing a lexical database of ASL (\ie ASL-LEX~\cite{caselli2017asl}).  & Unknown \\\hline
Signing Savvy~\cite{signingsavvy} & >13K & - & - & -  &Video, Gloss, Utterance, Word & Glosses include manual English-based labels. & Unknown \\\hline
How2Sign~\cite{duarte_how2sign_2021}& 16K & 80  &  11 & 1280 $\times$ 720 & Video, Pose, Gloss, Utterance, Speech & Glosses include English-based labels, but do not include information such as hand-shape, hand movement/orientation, and facial expressions, such as raised eyebrows in yes/no questions. & ELAN~\cite{crasborn2008enhanced} \\\hline
OpenASL~\cite{shi2022open} & 33K & 288 &  220 & - & Video, Utterance & - & - \\\hline
\bottomrule

\end{tabular}}
\end{table*}
\section{Gloss Annotation Conventions}
\small
\begin{longtable}{p{0.15\textwidth}|p{0.14\textwidth}|p{0.22\textwidth}|p{0.4\textwidth}}
    \toprule\hline
     \multicolumn{1}{c|}{\textbf{Category}} &  \multicolumn{1}{c|}{\textbf{Gloss}} &  \multicolumn{1}{c|}{\textbf{Example}} &  \multicolumn{1}{c}{\textbf{Explanation}} \\
    \hline
    \multirow{4}{*}{English-based glosses} & \multirow{2}{*}{-} & OH-I-SEE & \multirow{2}{6cm}{Used to separate words if the English translation of a single sign requires more than one.} \\\cline{3-3}
    &  & THANK-YOU &  \\\cline{2-4}
    & \multirow{2}{*}{/} & BOLD/TOUGH & \multirow{2}{6cm}{Used when one sign has two different English equivalents.} \\\cline{3-3}
    &  & THANK-YOU &  \\\hline
    \multirow{2}{*}{Fingerspelling} & fs- & fs-J-O-H-N & Fingerspelled word. \\\cline{2-4}
    & $\#$ & $\#$EARLY & Fingerspelled loan sign. \\\hline
    Name Signs & ns- & ns-PARIS & Used for names of places (\eg Paris). \\\hline
    \multirow{2}{*}{Compounds} & \multirow{2}{*}{+} & \multirow{2}{*}{MOTHER+FATHER} & A type of sign formation where two or more signs are joined to create a new sign with a distinct meaning (\eg ``parent''). \\\hline
    Phonological issues & QMwg & FRIEND FINISH DRIVE QMwg & Question marking sign (with wiggling) \\\hline
    \multirow{7}{2.5cm}{Subject and object verb agreement } & \multirow{3}{*}{i:GLOSS:j} & \multirow{2}{*}{i:GIVE:j} & ``i'' and ``j'' designate unique spatial locations associated with the subject and object referents. \\\cline{3-4}
    &  & 1p:GIVE:2p & ``(I) give (you)...''  \\\cline{2-4}
    & Noun & fs-J-O-H-N i:GIVE:j & John is signed in a neutral location. \\\cline{2-4}
    & \multirow{3}{*}{Noun:i} & \multirow{3}{*}{fs-J-O-H-N:i i:GIVE:j} & John is signed in the location associated with the referent (the same location with which the verb displays manual subject-verb agreement). \\\hline
    \multirow{6}{2.5cm}{Agreement marking on adjectives, nouns, pronouns, determiners, possessives, and emphatic reflexives} & \multirow{6}{*}{\makecell[l]{Pronoun\\ IX-[person]:i \\ \\ Determiner\\ IX-3p:i \\ }} & IX-1p & 1st person pronoun \\\cline{3-4}
    &  & POSS-1p & 1st person possessive marker \\\cline{3-4}
    &  & SELF-1p & 1st person emphatic reflexive marker (as in ``I did it myself'') \\\cline{3-4}
    &  & IX-2p & Pronoun referring to addressee. \\\cline{3-4}
    &  & POSS-2p & Possessive marker referring to addressee. \\\cline{3-4}
    &  & SELF-2p & Emphatic reflexive marker referring to addressee. \\\cline{3-4}
    & \multirow{6}{2.5cm}{\makecell[l]{Possessive\\ POSS-[person]:i \\ \\ Emphatic reflexive\\ SELF-[person]:i}} & \multirow{2}{6cm}{IX-3p:i} & Pronoun or determiner referring to singular third person referent associated with location ``i''. \\\cline{3-4}
    &  & \multirow{2}{6cm}{POSS-3p:i} & Possessive marker referring to singular third person referent associated with location ``i''. \\\cline{3-4}
    &  & \multirow{2}{6cm}{SELF:i} & Emphatic reflexive marker referring to a singular third person referent associated with location ``i''. \\\cline{2-4}
    & \multirow{2}{*}{-} & \multirow{2}{6cm}{THUMB-IX-3p:i} & Pronoun referring to singular third person referent associated with location ``i'' articulated with the thumb. \\\hline
    \multirow{6}{2.5cm}{Adverbials of location and direction} & \multirow{6}{2cm}{\makecell[l]{Adverbial \\ IX-loc:[location] \\ IX-dir:[direction]}} & {IX-loc:i} & Adverbial produced with index finger pointing to location ``i''. \\\cline{3-4}
    &  & IX-loc"under table" & \multirow{5}{*}{Adverbial with location described.} \\\cline{3-3}
    &  & IX-dir"around the corner to the right" &  \\\cline{3-3}
    &  & IX-loc"far" &  \\\cline{3-3}
    &  & THUMB-IX-loc"behind" &  \\\hline
    \multirow{13}{2.5cm}{Singular vs. plural}& \multirow{7}{2cm}{\makecell[l]{IX-[person]-[num]:i/j }} & IX-3p-pl-2:x/y & \multirow{2}{6cm}{Third person pronoun referring to the 2 (or 3) referents: x, y (or z).} \\\cline{3-3}
    &  & IX-3p-pl-3:x/y/z &  \\\cline{3-4}
    &  & \multirow{2}{2.9cm}{IX-1p-pl-2:x} &  First person pronoun referring to singer plus the referent associated with the location ``i''.\\\cline{3-4}
    &  & \multirow{2}{2.9cm}{IX-2p-pl-2:x} & Second person pronoun referring to addressee plus the two referents associated with locations ``x'' and ``y''. \\\cline{2-4}
    & \multirow{6}{2.2cm}{-3p-pl-arc} & IX-3p-pl-arc & \multirow{3}{6cm}{Pronoun (or possessive or emphatic reflexive) referring to singular third person referent associated with location ``i'' articulated with the thumb.} \\\cline{3-3}
    &  & POSS-3p-pl-arc &  \\\cline{3-3}
    &  & SELF-3p-pl-arc &  \\\cline{3-4}
    &  & \multirow{1}{2.9cm}{1p:GIVE-3p-arc} & \makecell[l]{``I give (it) to them.'' \\ Subject agreement is 1st person. Object agreement (the end \\ point of the sign) is plural (an arc).} \\\cline{2-4}
    & -loc-arc & IX-loc-arc & Adberbial (``there'') using an arc to designate locations. \\\hline
    \multirow{2}{2.5cm}{Reduplicative aspect marking}& Gloss-aspect & STUDY-continuative & Aspectual inflections are indicated following the gloss. \\\cline{2-4}
    & Gloss-aspect(:i) & GIFT-distributive:i & ``(they) each gave (one person)...'' \\\hline
    \multirow{2}{2.5cm}{Reciprocal inflection} & \multirow{2}{2cm}{GLOSS-recip} & \multirow{2}{6cm}{LOOK-AT-recip:i,j} & The referents associated with locations ``i'' and ``j'' look at each other. \\\hline
    \bottomrule
\end{longtable}\label{tab:gloss_convention}

\begin{table*}[htb!]
\footnotesize
\centering
\caption{Experimental results for different setups of English text-to-ASL gloss translation. Note: For ``Fine-tuning,'' the model was not constrained to the word-to-gloss dictionary vocabulary, unlike in few-shot prompting.
\Description[]}\label{tab:llm_experiment_results}
\renewcommand{\arraystretch}{1.1}
 \resizebox{\textwidth}{!}{
\begin{tabular}
{C{0.15\textwidth}|C{0.15\textwidth}|C{0.1\textwidth}|C{0.2\textwidth}|C{0.15\textwidth}}
\toprule\hline
\multicolumn{1}{c|}{{\textbf{\makecell[c]{Model}}}} &\multicolumn{1}{c|}{{\textbf{\makecell[c]{Training Method}}}} &\multicolumn{1}{c|}{{\textbf{\makecell[c]{Limited Vocab}}}} &\multicolumn{1}{c|}{{\textbf{\makecell[c]{Number of Examples}}}} & \multicolumn{1}{c}{{\textbf{\makecell[c]{BLEU-4 \bm{$\uparrow$}}}}}  \\\hline 
GPT-2 & Fine-tuning & - & 1474 (80\% of the entire dataset) & <0.000 \\\hline
\multirow{2}{*}{GPT-3.5-turbo-0125} & Few-shot prompting & No & 100 & 0.102\\\cline{2-5}
 & Fine-tuning & - & 1474 (80\% of the entire dataset) & 0.161 \\\hline
\multirow{2}{*}{GPT-4-turbo-2024-04-09} & \multirow{2}{*}{Few-shot prompting} & \multirow{2}{*}{No} & 100 & 0.115 \\\cline{4-5}
&  &  & 300 & 0.145 \\\hline
\multirow{3}{*}{GPT-4-0125-preview}& \multirow{3}{*}{Few-shot prompting} & \multirow{2}{*}{No} & 100 & 0.117 \\\cline{4-5}
&  &  & 300 & 0.143 \\\cline{3-5}
&  & Yes & 300 & 0.176 \\\hline
\multirow{3}{*}{\makecell[c]{GPT-4o-2024-05-13 \\ (Our adopted model)}} & \multirow{3}{*}{Few-shot prompting} & \multirow{2}{*}{No} & 100 & 0.133 \\\cline{4-5}
&  &  & 300 & 0.173 \\\cline{3-5}
&  & \textbf{Yes} & \textbf{300} & \textbf{0.226} \\\hline
 \bottomrule
\end{tabular}}
\end{table*}

\begin{table*}[htb!]
\footnotesize
\centering
\caption{Prompts for different setups.
\Description[]}\label{tab:prompt_engineering}
\renewcommand{\arraystretch}{1.1}
 \resizebox{\textwidth}{!}{
\begin{tabular}
{C{0.1\textwidth}|C{0.1\textwidth}|L{0.6\textwidth}}
\toprule\hline
\multicolumn{1}{c|}{{\textbf{\makecell[c]{Limited Vocab}}}} &\multicolumn{1}{c|}{{\textbf{\makecell[c]{Grammar Rules}}}} &\multicolumn{1}{c}{{\textbf{\makecell[c]{Prompts}}}}\\\hline 
 \multirow{3}{*}{No} & No & You are an ASL translator. Your task is to translate an English sentence to an ASL gloss format. \\\cline{2-3}
  & Yes & You are an ASL translator. Your task is to translate an English sentence to an ASL gloss format. First, familiarize yourself with the following ASL grammar rules: \textcolor{RoyalBlue}{\textsf{GRAMMER$\_$RULES}}. \\\hline
 \multirow{5}{*}{Yes} & No & You are an ASL translator. Your task is to translate an English sentence into ASL gloss format. First, familiarize yourself the following vocabulary dictionary: \textcolor{RoyalBlue}{\textsf{TEXT$\_$TO$\_$GLOSS$\_$DICTIONARY}}.  \\\cline{2-3}
  & Yes & You are an ASL translator. Your task is to translate an English sentence into ASL gloss format. First, familiarize yourself with the following ASL grammar rules: \textcolor{RoyalBlue}{\textsf{GRAMMER$\_$RULES}}. Also, review the following vocabulary dictionary: \textcolor{RoyalBlue}{\textsf{TEXT$\_$TO$\_$GLOSS$\_$DICTIONARY}}.  \\\hline
 \bottomrule
\end{tabular}}
\end{table*}

\begin{table*}[htb!]
\caption{Evaluation results of translating English text into glosses (Task on the left side in Module 1) using RAG. \bm{$^*$}All BLEU-4 and SacreBLEU scores are identical. \bm{$\uparrow$} indicates that higher values represent better performance, while \bm{$\downarrow$} indicates that lower values represent better performance. Best results in \textbf{bold}. The presented results are $mean\pm std$ across 10 repetitions of test set evaluation. Note: If ``Anonymized Embeddings'' is set to ``No'', RAG was performed using embeddings of the original data, else, it was performed using embeddings of the anonymized data.
\Description[This table presents evaluation results of translating English text to English-based glosses. The first row contains ten headers, including number of examples, anonymized embeddings, bleu-1, bleu-2, bleu-3, bleu-4, rouge-l, meteor, chrf, and ter. Regarding the columns, the first column shows number of used examples, where the first row is for using 1474 (all examples -not using RAG) and the other rows are for using 200, 100, and 50 examples. The second column, anonymized embeddings, indicates whether anonymized embeddings are used (yes or no). The remaining columns display various evaluation metrics with the symbol uparrow indicating higher values are better, and downarrow indicating lower values are better. Key findings are: presenting the model with less examples that are more relevant yields better results than using all examples. Moreover, in most cases using less examples (n=50) gives better results than using more examples (e.g. 100, 200). However, for some metrics (ROUGE, TER) using more examples gives better results. The best scores are bolded across all metrics, indicating the optimal settings for this translation model.]}\label{tab:text-to-gloss_RAG_eval_results_appx} 

\renewcommand{\arraystretch}{1.1}
 \resizebox{\textwidth}{!}{
\begin{tabular}{c | c | c c c c c c c c }
\toprule\hline
\multicolumn{1}{c|}{{\textbf{\makecell[c]{Number of \\ Examples}}}} &\multicolumn{1}{c|}{{\textbf{\makecell[c]{Anonymized \\ Embeddings}}}} & \multicolumn{1}{c}{{\textbf{\makecell[c]{BLEU-1 \bm{$\uparrow$}}}}} &\multicolumn{1}{c}{{\textbf{\makecell[c]{BLEU-2 \bm{$\uparrow$}}}}} & \multicolumn{1}{c}{{\textbf{\makecell[c]{BLEU-3 \bm{$\uparrow$}}}}}
& \multicolumn{1}{c}{{\textbf{\makecell[c]{BLEU-4\bm{$^*$} \bm{$\uparrow$}}}}} 
& \multicolumn{1}{c}{{\textbf{\makecell[c]{ROUGE-L \bm{$\uparrow$}}}}} 
& \multicolumn{1}{c}{{\textbf{\makecell[c]{METEOR \bm{$\uparrow$}}}}} 
& \multicolumn{1}{c}{{\textbf{\makecell[c]{CHrF \bm{$\uparrow$}}}}} & \multicolumn{1}{c}{{\textbf{\makecell[c]{TER \bm{$\downarrow$}}}}} \\\hline 
 \multirow{2}{*}{\makecell[c]{200}} & No & $0.562\pm0.003$ & $0.433\pm0.003$ & $0.345\pm0.003$ & $0.278\pm0.003$ & \bm{$0.669\pm0.001$} & $0.56\pm0.001$ & $0.557\pm0.002$ & \bm{$0.52\pm0.003$}\\
 & Yes & \bm{$0.569\pm0.003$} & \bm{$0.437\pm0.004$} & $0.347\pm0.004$ & $0.278\pm0.003$ & $0.668\pm0.002$ & $0.559\pm0.006$ & \bm{$0.559\pm0.001$} & \bm{$0.52\pm0.002$} \\ \hline
 \multirow{2}{*}{\makecell[c]{100}} & No & $0.563\pm0.003$ & $0.433\pm0.003$ & $0.345\pm0.002$ & $0.278\pm0.003$ & $0.663\pm0.003$ & $0.556\pm0.003$ & $0.557\pm0.001$ & $0.522\pm0.004$\\
 & Yes & $0.567\pm0.003$ & \bm{$0.437\pm0.003$} & $0.345\pm0.002$ & \bm{$0.279\pm0.003$} & $0.666\pm0.002$ & $0.562\pm0.002$ & \bm{$0.559\pm0.002$} & $0.523\pm0.002$\\ \hline
 \multirow{2}{*}{\makecell[c]{50}} & No & $0.563\pm0.003$ & $0.432\pm0.003$ & $0.342\pm0.004$ & $0.275\pm0.005$ &  $0.663\pm0.002$ & $0.557\pm0.003$ & $0.554\pm0.003$ & $0.525\pm0.006$ \\
 & Yes & \bm{$0.569\pm0.003$} & \bm{$0.437\pm0.003$} & \bm{$0.348\pm0.003$} & \bm{$0.279\pm0.003$} & $0.667\pm0.002$ & \bm{$0.564\pm0.002$} & $0.558\pm0.002$ & $0.523\pm0.002$ \\ \hline 
 \bottomrule
\end{tabular}
}
\end{table*}

\begin{table}[htb!]
    \small
    \centering
    \caption{Existing English Text-to-ASL Gloss translation results reported in the literature.}
    \begin{tabular}{l|l|r}
    \toprule\hline
        \makecell[c]{\textbf{References}} & \makecell[c]{\textbf{Dataset}} & \textbf{BLEU-4}\\\hline
         Inan \etal~\cite{inan2024generating} &  Self-Collected ASL Dataset &0.002\\
        Zhu \etal~\cite{zhu_neural_2023} & NCSLGR & 0.124 \\
        Moryossef \etal~\cite{moryossef_data_2021} & NCSLGR & 0.191\\\hline
         \bottomrule
    \end{tabular}
    \label{tab:text-to-gloss_sota}
\end{table}

\end{document}